\documentclass[superscriptaddress,aps,pra,twocolumn,10pt,footinbib]{revtex4-1}
\usepackage[T1]{fontenc}
\usepackage{amssymb,amsthm,amsmath,amsfonts}
\usepackage{xcolor,graphicx}
\usepackage[colorlinks=true,allcolors=blue]{hyperref}
\usepackage[caption=false,labelformat=empty,subrefformat=parens]{subfig}
\usepackage{sistyle}
\usepackage{lipsum}
\usepackage{mathtools}

\begin{document}

\title{Heat transport in carbon nanotubes: Length dependence of phononic conductivity from the Boltzmann transport equation and molecular dynamics}
\author{Daniel Bruns}
\affiliation{Department of Physics and Astronomy, University of British Columbia, Vancouver, BC, Canada V6T 1Z1}
\affiliation{Quantum Matter Institute, University of British Columbia, Vancouver, BC, Canada V6T 1Z4}
\author{Alireza Nojeh}
\affiliation{Department of Electrical and Computer Engineering, University of British Columbia, Vancouver, BC, Canada V6T 1Z4}
\affiliation{Quantum Matter Institute, University of British Columbia, Vancouver, BC, Canada V6T 1Z4}
\author{A. Srikantha Phani}
\affiliation{Department of Mechanical Engineering, University of British Columbia, Vancouver, BC, Canada V6T 1Z4}
\author{J\"org Rottler}
\affiliation{Department of Physics and Astronomy, University of British Columbia, Vancouver, BC, Canada V6T 1Z1}
\affiliation{Quantum Matter Institute, University of British Columbia, Vancouver, BC, Canada V6T 1Z4}

\date{\today}

\begin{abstract}

In this article, we address lattice heat transport in single-walled carbon nanotubes (CNTs) by a quantum mechanical calculation of three-phonon scattering rates in the framework of the Boltzmann transport equation (BTE) and classical molecular dynamics (MD) simulation.
Under a consistent choice of an empirical, realistic atomic interaction potential, we compare the tube length dependence of the lattice thermal conductivity (TC) at room temperature determined from an iterative solution of the BTE and from a nonequilibrium MD (NEMD) approach.
Qualitatively similar trends are found in the limit of short tubes, where an extensive regime of ballistic heat transport prevailing in CNTs of lengths $L\lesssim 1\,\rm{\mu m}$ is independently confirmed. In the limit of long tubes, the BTE approach suggests a saturation of TC with tube length, whereas direct NEMD simulations of tubes extending up to $L=10\,\rm{\mu m}$ are demonstrated to be insufficient to settle the question of whether a fully diffusive heat transport regime and an intrinsic value of TC exist for CNTs. Noting that acoustic phonon lifetimes lie at the heart of a saturation of TC with tube length as per the BTE framework, we complement the quantum mechanical prediction of acoustic phonon lifetimes with an analysis of phonon modes in the framework of equilibrium MD (EMD). A normal mode analysis (NMA) with an emphasis on long wavelength acoustic modes corroborates the BTE prediction that heat transport in CNTs in the long tube limit is governed by the low attenuation rates of longitudinal and twisting phonons.
\end{abstract}

\maketitle

\section{Introduction}

Carbon nanotubes (CNTs) are commonly suggested as a test ground for the unusual thermal transport properties that manifest in low dimensional lattice models~\cite{D08}.
The most prominent anomaly of heat conduction in low dimensions  is the domain size dependence of lattice thermal conductivity (TC) $\kappa$. 
For some strictly one-dimensional models of nonlinear chains, a power law divergence of TC with system size $L$ has been predicted, $\kappa \sim L^\alpha$ where $0<\alpha<1$, and linked to the presence of weakly damped long wavelength vibrational modes~\cite{LLP03,LLP16}. 
Advancing from model systems to real physical materials with reduced dimensionality, numerous studies, both experimental~\cite{COG08,LLH17,LWL17,LTZ17} and theoretical~\cite{M02,ZL05,MB05,MBPRL05,DG07,SHK09,LBM09,TIM10,CQ12,SOV15,FDH19}, have aimed at unveiling the tube length dependence of CNT thermal conductivity. In CNTs, heat is carried predominantly by phonons~\cite{HWP99,YWW04,B11} and, considering pristine tubes, intrinsic resistance to phonon transport arises exclusively from lattice anharmonicity. In comparison to other materials, phonons in CNTs exhibit a relatively long intrinsic mean free path and it is generally acknowledged that TC becomes a length dependent quantity at the submicron scale~\cite{MPG13} where physical boundaries provide a major contribution to thermal resistance~\cite{CBG14}. In the limit of long tubes, however, conflicting predictions persist as to whether TC saturates to a finite, length independent value~\cite{LLH17,MB05,LBM09,MBPRL05,DG07,TIM10,FDH19} as is the case for diffusive (Fourier) transport or diverges with tube length~\cite{COG08,LWL17,M02,ZL05,SHK09,CQ12} such that heat transport is in fact super-diffusive.

Existing theoretical predictions of the phononic TC as a function of tube length stem either from an explicit quantum mechanical calculation of phonon scattering rates in the framework of the Boltzmann transport equation (BTE)~\cite{MB05,MBPRL05,LBM09} or from classical molecular dynamics (MD) simulations~\cite{M02,ZL05,DG07,SHK09,TIM10,CQ12,SOV15,FDH19}. In determining the resistance to heat flow in pristine crystals, the BTE approach relies on a perturbative treatment of lattice anharmonicity and calculations typically truncate scattering processes with more than three phonons involved. Mingo et al.~\cite{MB05} argued that this limitation causes TC predictions to diverge in the limit of long tubes, since it is higher order phonon-phonon scattering that ensures the dissipation of heat carried by long wavelength acoustic phonons. Treating higher order processes approximately, BTE calculations by Lindsay et al.\ indicated that millimeter long tubes are required to observe fully diffusive transport in CNTs~\cite{LBM09}. Trends of TC as a function of system size derived from MD, on the other hand, naturally include all orders of phonon-phonon interaction. Still, based on MD, some studies reported a power law divergence of TC~\cite{M02,ZL05,SHK09,CQ12}, whereas others found convergence with tube length~\cite{DG07,TIM10,FDH19}. Arguably, the use of classical phonon statistics in MD introduces significant error into TC predictions when applied to materials like CNTs that possess a high Debye temperature~\cite{MBPRL05}. Yet general trends of TC with tube length, in particular the emergence of a diffusive transport regime, should remain unaffected by the choice of quantum or classical phonon statistics in either computational framework. 

In an effort to shed light on the discrepant theoretical predictions, this article revisits the length dependence of phonon mediated heat transport in CNTs by comparing trends that ensue from both the BTE and the MD framework under a consistent choice of atomic interaction potential. We employ an iterative solution to the linearized BTE~\cite{OS95,BWM05,FLP13} which, in addition to anharmonic three-phonon processes, incorporates the finite tube length as a source of resistive phonon scattering~\cite{MB05,LBM09,LBM10}. By simultaneously conducting nonequilibrium molecular dynamics (NEMD) simulation of CNTs of varying lengths, we contrast the transition of heat transport away from the ballistic regime as predicted by both computational approaches for tubes of distinct diameters. Through NEMD simulations, we show that TC values continue to increase even if the distance between heat source and sink is extended up to $L=10\,\rm{\mu m}$ to the effect that no conclusions about a saturation with tube length can be drawn. Bypassing the computational limitations of the NEMD approach, the large $L$ limit of heat transport is addressed in the BTE framework. Here, we demonstrate that a fully diffusive heat transport regime can be established and traced back to the three-phonon scattering rates of acoustic phonons in the long wavelength limit.
 To further elucidate the role of acoustic phonons in determining the length dependence of TC and to allow for a possible impact of higher order phonon-phonon scattering, we subsequently perform normal mode analysis (NMA) in the framework of equilibrium molecular dynamics~\cite{LMH86,MK04,K09,TTI10}. In contrast to earlier work that applied NMA to study thermal transport properties of CNTs~\cite{DG07,TTI10,OPS11}, we exclusively consider long wavelength acoustic modes. By omitting high frequency phonon modes and by calculating acoustic normal mode coordinates on-the-fly, we are able to resolve modes of unprecedented wavelength in tubes of varying diameters. The so-detected phonon lifetimes extend into the nanosecond range, which substantiates the decisive role of emerging long wavelength acoustic modes in governing the length dependence of TC.  

This article is structured as follows. In Sec.~\ref{sec:FRMWRKS}, we provide a general description of the computational frameworks used to study phonon mediated heat transfer in crystals. For CNTs, details specific to the calculation of TC employed in this work are given in Sec.~\ref{sec:CMPT_DTLS}. We discuss our results in Sec.~\ref{sec:RSLTS_DIS} and conclude in Sec.~\ref{sec:CNCLS}.

%

\section{Frameworks to calculate phononic thermal conductivity}\label{sec:FRMWRKS}

\subsection{Boltzmann transport equation}

In the framework of the Boltzmann transport equation (BTE), lattice vibrations of a periodic crystal are conceived as a kinetic gas of phonons. The BTE for phonons states that a temperature gradient leads to a drift motion of phonons, which is balanced by resistive particle-like scattering processes~\cite{P29,Z60,K69}. This notion allows one to express the contributions of individual phonons to the conductivity as
\begin{align}\label{eq:BTE}
\kappa_\lambda^{x}=c_\lambda v_{x,\lambda}^2\tau_\lambda,
\end{align}
where the label $\lambda=(\boldsymbol{q},j)$ subsumes the wave vector $\boldsymbol{q}$ and polarization dependence $j$ of each phonon. Mode-specific conductivity contributions follow from the volumetric heat capacity $c_\lambda=1/V(\hbar \omega_{\lambda}\partial n_{\lambda}/\partial{T})$ with ${n_{\lambda}=[\exp{(\hbar \omega_{\lambda}/k_{{\rm{B}}}T)}-1]^{-1}}$ standing for the Bose-Einstein statistics, the phonon group velocity in transport direction $v_{x,\lambda}=\partial \omega_{\lambda}/ \partial q_x$, and the lifetime $\tau_{\lambda}$ which constitutes a measure of resistive phonon scattering. Phonon heat capacities $c_{\lambda}$ and group velocities $v_{x,\lambda}$ are derived from a harmonic lattice calculation that requires second order force constants $\Phi_{lb,l'b'}^{\alpha,\alpha'}$ and a diagonalization of the dynamical matrix 
\begin{align}\label{eq:dynMat}
D^{\alpha,\alpha'}_{b,b'}(\boldsymbol{q})=\frac{1}{\sqrt{m_b m_{b'}}}\sum_{l'}\Phi_{0b,l'b'}^{\alpha,\alpha'}\,e^{i \boldsymbol{q}\cdot \boldsymbol{r}_{l'}}
\end{align}
whose eigenvalues, $D(\boldsymbol{q})\boldsymbol{\rm{e}}^{\lambda}=\omega_{\lambda}^{2}\boldsymbol{\rm{e}}^{\lambda}$, yield the harmonic phonon spectrum, $\omega_\lambda=\omega_j(\boldsymbol{q})$.
In the above expression, superscripts run over the three spatial dimensions and $\boldsymbol{r}_{l'}$ represents the position vector of the $l'$th unit cell whose atoms of mass $m_b$ reside at $\boldsymbol{r}_{l',b}=\boldsymbol{r}_{l'}+\boldsymbol{r}_{b}$.

To the extent that anharmonic terms of the interatomic interaction potential can be regarded as a weak perturbation of harmonic phonon modes, finite phonon lifetimes resulting from anharmonic three-phonon scattering processes can be calculated in lowest order perturbation theory~\cite{MSB14,FR14} as
\begin{align}\label{eq:LTIMES_3rd_RTA}
\frac{1}{\tau_{3,\lambda}^0}=\frac{1}{N}\left(\sum_{\lambda',\lambda''}^{-}\frac{1}{2}\Gamma^{-}_{\lambda,\lambda',\lambda''}
+\sum_{\lambda',\lambda''}^{+}\Gamma^{+}_{\lambda,\lambda',\lambda''}\right),
\end{align}
with $N$ denoting the number of translational unit cells and where the first and second sum extend over phonon splitting, $\lambda\rightarrow \lambda' +\lambda''$, and absorption, $\lambda+\lambda'\rightarrow \lambda''$, processes with individual transition amplitudes~\cite{MSB14,FR14} given as 
\begin{align}\label{eq:3rd_ampltudes}
\Gamma^{\pm}_{\lambda,\lambda',\lambda''}=\frac{\hbar \pi}{4\omega_{\lambda}\omega_{\lambda'}\omega_{\lambda''}}\begin{Bmatrix}n_{\lambda'}-n_{\lambda''}\\
1+n_{\lambda'}+n_{\lambda''}
\end{Bmatrix}\left|V^{\pm}_{\lambda,\lambda',\lambda''}\right|^2.
\end{align}
Here, the top (bottom) row of the curly brackets corresponds to absorption (splitting) processes and transition matrix elements depend on second $\Phi_{lb,l'b'}^{\alpha,\alpha'}$ and third $\Phi_{lb,l'b',l''b''}^{\alpha,\alpha',\alpha''}$ order force constants~\cite{MSB14,FR14}, 
\begin{align}\label{eq:3rd_matrix}
V^{\pm}_{\lambda,\lambda',\lambda''}=\hspace{-0.29cm}\sum_{\substack{\alpha,\alpha',\alpha''\\b,l'b',l''b''}}\hspace{-0.25cm}\Phi_{0b,l'b',l''b''}^{\alpha,\alpha',\alpha''}\frac{{\rm{e}}^{\lambda}_{\alpha b}{\rm{e}}^{\pm\lambda'}_{\alpha' b'}{\rm{e}}^{-\lambda''}_{\alpha'' b''}}{\sqrt{m_b m_{b'}m_{b''}}}e^{\pm i\boldsymbol{q'}\cdot\boldsymbol{r}_{l'}}e^{- i\boldsymbol{q''}\cdot\boldsymbol{r}_{l''}},
\end{align}
where $\boldsymbol{\rm{e}}^\lambda$ are the harmonic phonon eigenvectors of the dynamical matrix~\eqref{eq:dynMat} and $-\lambda\equiv (-\boldsymbol{q},j)$. In Eq.~\eqref{eq:LTIMES_3rd_RTA}, the summation symbols are marked with a plus and minus sign, $\sum^{\pm}$, which signifies a restriction to three-phonon processes that obey both the energy, $\omega_{\lambda}\pm\omega_{\lambda'}=\omega_{\lambda''}$, and the momentum, $\boldsymbol{q} \pm \boldsymbol{q}'=\boldsymbol{q}''+\boldsymbol{G}$, selection rules with $\boldsymbol{G}$ denoting a reciprocal lattice vector. $\boldsymbol{G}=0$ refers to normal and $\boldsymbol{G}\neq 0$ to umklapp processes.

The superscript of lifetimes $\tau^{0}_{3,\lambda}$ computed by means of Eqs.~\eqref{eq:LTIMES_3rd_RTA}-\eqref{eq:3rd_matrix} indicates a so-called relaxation time approximation (RTA). Plugging the RTA lifetimes $\tau^{0}_{3,\lambda}$ into Eq.~\eqref{eq:BTE} constitutes an approximation to TC inasmuch as assuming that both normal and umklapp processes contribute equally to the dissipation of heat inside a crystal, while in fact normal processes contribute only indirectly by redistributing the per-phonon mode energy~\cite{Z60}. Derived from a formally exact solution of the linearized BTE for phonons~\cite{OS95,BWM05,MSB14,FR14}, a more accurate description of lifetimes appearing in Eq.~\eqref{eq:BTE} is given by
\begin{align}\label{eq:iterBTE}
\tau_\lambda=\tau_\lambda^0(1+\Delta_\lambda),
\end{align}
where $\tau^0_{\lambda}$ represents an RTA solution to TC and the interplay of normal and umklapp scattering causing thermal resistance is reflected by the term $\Delta_{\lambda}$ which, if only anharmonic three-phonon processes in lowest order perturbation theory are considered, takes the form
\begin{equation}
\begin{aligned}\label{eq:Delta_3}
\Delta_{\lambda}=&\frac{1}{N}\sum^-_{\lambda',\lambda''}\frac{1}{2}\Gamma^{-}_{\lambda,\lambda',\lambda''}(\xi_{\lambda,\lambda''}\tau_{\lambda''}+\xi_{\lambda,\lambda'}\tau_{\lambda'})\\
&+\frac{1}{N}\sum^+_{\lambda',\lambda''}\Gamma^{+}_{\lambda,\lambda',\lambda''}(\xi_{\lambda,\lambda''}\tau_{\lambda''}-\xi_{\lambda,\lambda'}\tau_{\lambda'}),
\end{aligned}
\end{equation}
with the shorthand notation $\xi_{\lambda,\lambda'}=\omega_{\lambda'}v_{x,\lambda'}/\omega_{\lambda}v_{x,\lambda}$ and scattering amplitudes defined by Eq.~\eqref{eq:3rd_ampltudes}. The unknown lifetimes $\tau_{\lambda}$ of each mode are coupled through the term $\Delta_{\lambda}$ such that a solution to Eq.~\eqref{eq:iterBTE} may be found by carrying out an iteration $\tau_\lambda^{(k+1)}=\tau_{\lambda}^{0}(1+\Delta_{\lambda}^{(k)})$ with $\Delta^{(0)}_{\lambda}$ initially calculated based on RTA lifetimes $\tau^{0}_{\lambda}$ .

Starting from Eq.~\eqref{eq:LTIMES_3rd_RTA}, additional phonon scattering rates stemming from processes such as phonon-isotope $(\tau^{0}_{{\rm{iso}},\lambda})^{-1}$ or phonon-boundary scattering $(\tau^{0}_{{\rm{bs}},\lambda})^{-1}$ can be included into the BTE framework by expanding RTA lifetimes according to Matthiessen's rule as 
\begin{align}
\frac{1}{\tau_{\lambda}^{0}}=\frac{1}{\tau^0_{3,\lambda}}+\frac{1}{\tau^0_{{\rm{iso}},\lambda}}+\frac{1}{\tau^0_{\rm{bs},\lambda}}+\dots
\end{align}
and, if the dissipative vs.\ nondissipative character of the added scattering channels is considered, by further extending the term $\Delta_{\lambda}$~\cite{MSB14,FR14}.

\subsection{Nonequilibrium Molecular Dynamics}

Heat transport through a crystal can be probed directly by means of molecular dynamics (MD). With the goal of determining a crystal's TC, the nonequilibrium molecular dynamics (NEMD) method~\cite{CT80,IH94,MP97,JJ99,MBC15} consists in the simulation of a steady-state heat transport regime during which the very definition of TC from Fourier's law is applied,
\begin{align}\label{eq:FL_NEMD}
\kappa^x=-\frac{J^x}{\nabla T},
\end{align}
where $J^x$ is the heat flux density along the direction of the temperature gradient $\nabla T$.

Within the framework of MD, a nonequilibrium transport regime can be induced, for example, by directly thermostatting a heat sink and a source at different temperatures~\cite{MBC15} or, indirectly, by continuously swapping atomic velocity vectors between two predetermined hot and cold regions of the simulation domain~\cite{MP97}.
To measure a temperature profile, the simulation domain is divided into $M$ slabs along the transport direction, $\{x_1,\dots,x_M\}$. While observing the temperature of each slab, $\{T(x_1),\dots,T(x_M)\}$, the system is propagated in time until the steady-state transport regime is established. Finally, a gradient $\nabla T$ can be inferred from the temperature profile which, in conjunction with the tallied heat flux $J^x$, enables the calculation of TC by means of Eq.~\eqref{eq:FL_NEMD}.

It is worth mentioning that NEMD derived temperature profiles often exhibit nonlinearities in the proximity of thermostatted regions. Typically observed abrupt jumps of temperature occurring at the boundaries of the hot and cold reservoirs leave some ambiguity as to the definition of the temperature gradient $\nabla T$. This ambiguity in the NEMD method was recently addressed by the work of Li et al.~\cite{LXS19}, showing that fitting only the linear part of the temperature profile leads to an overestimation of TC as calculated by Eq.~\eqref{eq:FL_NEMD}. Instead, conclusive evidence was given in~\cite{LXS19} that the NEMD method requires to compute  $\nabla T =\Delta T/L$, where $\Delta T$ is the temperature difference between the heat source and sink which are separated by a distance $L$.

\subsection{Normal Mode Analysis}

Another approach to TC at the level of individual phonon contributions involves an analysis of phonons in their representation as vibrational waves describing the decoupled normal modes of a harmonic crystal. 
The normal mode coordinates of a crystal with $N$ unit cells, 
\begin{align}\label{eq:Qnorm_coords}
Q_{\lambda}(t)&=\frac{1}{\sqrt{N}}\sum_{l}{\boldsymbol{\rm{e}}}^{\lambda *}\cdot \tilde{\boldsymbol{u}}_{l}(t)\,e^{i\boldsymbol{q}\cdot \boldsymbol{r}_{l}},\\\label{eq:Qdotnorm_coords}
\dot{Q}_{\lambda}(t)&=\frac{1}{\sqrt{N}}\sum_{l}{\boldsymbol{\rm{e}}}^{\lambda *}\cdot \tilde{\boldsymbol{v}}_{l}(t)\,e^{i\boldsymbol{q}\cdot \boldsymbol{r}_{l}},
\end{align}
can be calculated by projecting mass-weighted atomic displacements $\tilde{\boldsymbol{u}}_{l,b}=\sqrt{m_b}\boldsymbol{u}_{l,b}$ and velocities $\tilde{\boldsymbol{v}}_{l,b}=\sqrt{m_b}\boldsymbol{v}_{l,b}$ onto complex conjugated phonon eigenvectors $\boldsymbol{\rm{e}}^{\lambda *}$~\cite{D93}. In equilibrium molecular dynamics (EMD), atomic phase space coordinates are propagated in time under the influence of realistic interatomic potentials whose anharmonic terms cause the harmonic mode energy,
\begin{align}\label{eq:harm_E}
E_{\lambda}(t)=\frac{1}{2}\dot{Q}_\lambda(t)\dot{Q}^*_{\lambda}(t)+\frac{\omega^2_{\lambda}}{2}Q_{\lambda}(t)Q^*_{\lambda}(t),
\end{align}
to fluctuate with a mode-specific time scale $\tau^{\rm{MD}}_{\lambda}$. Adopting the BTE description of TC in terms of Eq.~(\ref{eq:BTE}), the program of normal mode analysis (NMA) estimates intrinsic phonon-specific thermal resistance by setting $\tau_{\lambda}=\tau_{\lambda}^{\rm{MD}}$~\cite{LMH86,MK04,K09,TTI10}.

Given a time series of normal mode coordinates obtained by means of equilibrium MD through Eqs.~\eqref{eq:Qnorm_coords} and~\eqref{eq:Qdotnorm_coords}, the time scale of energy fluctuations $\tau_{\lambda}^{\rm{MD}}$ can be determined, for example, by fitting an exponential decay to the normalized autocorrelation function of the harmonic per-mode energy~\cite{MK04},
\begin{align}\label{eq:ACF_E}
\frac{\langle E_{\lambda}(0)E_{\lambda}(t)\rangle}{\langle E_{\lambda}(0)E_{\lambda}(0)\rangle}=e^{-t/\tau_{\lambda}^{\rm{MD}}},
\end{align}
or, alternatively, by fitting the spectral density of the kinetic mode energy to a Lorentzian line shape~\cite{K09,TTI10,FR14}, 
\begin{align}\label{eq:KIN_E}
K_{\lambda}(\omega)=\frac{C_{\lambda}}{\left(\omega-\omega_{\lambda}\right)^{2}+\left(2\,\tau_{\lambda}^{\rm{MD}}\right)^{-2}},
\end{align}
where $K_{\lambda}(\omega)=|\mathcal{F}[\dot{Q}_{\lambda}(t)]|^{2}$ implies a Fourier transformation and $C_{\lambda}$ as well as $\omega_{\lambda}$ are left as additional fitting parameters. The latter approach requires input only in the form of atomic velocities which, however, comes at the cost of relying on a nonlinear fitting routine. 

\section{Computational Details}\label{sec:CMPT_DTLS}

Throughout this work, we consider armchair CNTs of chirality $(n,n)$ with diameter $D=(\sqrt{3} a/\pi)\, n$, where the translational lattice constant along the tube is $a\approx 2.5\,$\AA. For the cross-sectional area, we adopt the prevalent notion of an annular ring $A=\pi D\delta$ with $\delta=3.35\,$\AA~chosen to be the interlayer distance in graphite. The interatomic interaction potential and related force constants are assumed to be given by a Tersoff potential~\cite{TPRL88,T88} with an optimized parameterization for ${\rm{sp}}^2$-bonded carbon~\cite{LB10}. 

Calculations in the BTE framework are performed with the software package \textsl{shengBTE}~\cite{LCK14}. To calculate RTA lifetimes as stated in Eqs.~\eqref{eq:LTIMES_3rd_RTA}-\eqref{eq:3rd_matrix}, phonons are assumed to take on wavenumbers $q=(2\pi/Na)m$ with integers $-N/2<m\leq N/2$ implying a discretization of the CNT's first Brillouin zone with $N$ equidistant wavenumbers. In order to address the truncation of energy conserving three-phonon processes not sampled by the imposed $q$-grid, we adopt the default Gaussian smearing scheme implemented in \textsl{shengBTE} which aims to approximate transition amplitudes in the limit of continuous wavenumbers. Furthermore, in accordance with earlier work on the length dependence of TC~\cite{MB05,LBM09}, we extend the program to take into account a tube length dependent boundary scattering rate, ${(\tau_{{\rm{bs}},\lambda}^0)^{-1}=2|v_\lambda|/L}$, so that final RTA lifetimes are computed as ${(\tau_{\lambda}^0)^{-1}=(\tau_{3,\lambda}^0)^{-1}+(\tau_{{\rm{bs}},\lambda}^0)^{-1}}$. This choice of $(\tau^0_{\rm{bs},\lambda})^{-1}$ was shown to correctly reproduce the solution of the BTE in the limit of ballistic heat conduction, where the tube length $L\rightarrow 0$ and three-phonon scattering becomes negligible~\cite{MB05}. With initial values given by RTA lifetimes, the self-consistent phonon lifetimes defined in Eqs.~\eqref{eq:iterBTE}-\eqref{eq:Delta_3} are determined iteratively until, according to Eq.~\eqref{eq:BTE}, convergence of phononic TC, $\kappa^{x}=\sum_{\lambda}\kappa_{\lambda}^{x}$, with ${|\kappa^{x,(k+1)}-\kappa^{x,(k)}|<10^{-4}}\,\rm{Wm^{-1}K^{-1}}$ is achieved.

MD simulations are carried out using the \textsl{LAMMPS} package~\cite{P95}. We apply periodic boundary conditions along the tube axis in both NEMD and EMD simulations. Each simulation is preceded by an equilibration in the isothermal-isobaric ensemble $(T=300\,{\rm{K}},\,P=0)$ and by a subsequent equilibration in the microcanonical ensemble. 
 
To calculate TC by use of NEMD, we follow the velocity swapping protocol~\cite{MP97}. For tubes of varying size, we choose the frequency of velocity swaps such that the temperature difference between the hot and cold slabs remains within $\Delta T < 50\,\rm{K}$. Following Li et al.~\cite{LXS19}, we set $\nabla T=\Delta T/L$ to compute $\kappa^x=(\Delta E_{\rm{kin}}/2A\Delta t)/(\Delta T/L)$, where $\Delta E_{\rm{kin}}$ denotes the accumulated energy of velocity swaps over a steady state simulation runtime $\Delta t$, while the factor of 1/2 accounts for the heat flow in two directions under the imposed periodic boundary conditions along the tube. To address statistical uncertainties in the NEMD approach, for each tube size, we average TC values obtained from three independent simulations, where each simulation is performed with a slightly varied velocity swap frequency. More details on the NEMD analysis are given in Sec.~I of the Supplemental Material~\footnote{See Supplemental Material for details on the NEMD analysis, numerical artifacts stemming from flexural acoustic modes encountered by iteratively solving Eq.~\eqref{eq:iterBTE}, and spectral energy densities of acoustic normal mode coordinates.}.


Conducting NMA in combination with EMD simulations, we infer phonon lifetimes from atomic velocities in the frequency domain by means of Eqs.~\eqref{eq:Qdotnorm_coords} and~\eqref{eq:KIN_E}. The standard NMA protocol considers the full phonon spectrum which, in the case of CNTs, requires to sample atomic velocities with a period of ${T_{\rm{s}}< 10\,\rm{fs}}$ in order to resolve the highest phonon frequencies which extend up to ${\omega_{\lambda}^{\rm{max}}/2\pi\approx 51\,\rm{THz}}$. At the same time, normal mode coordinates have to be calculated and stored sequentially  $N_{Q}$ times to obtain a discrete Fourier transformation with frequency resolution, $\Delta \omega/2\pi=1/(N_{Q}T_{\rm{s}})$, that allows to capture the slow dynamics of long wavelength acoustic phonons. Here, we restrict NMA exclusively to acoustic modes with $\omega_{\lambda}/2\pi<2\,\rm{THz}$, which significantly reduces the required sampling frequency and the overall number of normal mode coordinates to consider. Specifically, while calculating normal mode coordinates on-the-fly during EMD runtime, we set $T_{\rm{s}}=256\,\rm{fs}$ to sample $N_Q=2^{18}$ consecutive dumps giving rise to a frequency resolution of $\Delta \omega/2\pi\approx 15\,\rm{MHz}$. Moreover, to reduce the noise level, we average the frequency spectra obtained from 40 independent EMD simulations, where each simulation is made independent by setting a different initial velocity seed. Finally, to determine acoustic phonon lifetimes as per Eq.~\eqref{eq:KIN_E}, we initialize a nonlinear least-square fitting routine with starting parameters deduced from the height, position and full width at half maximum of the most prominent peak in the frequency spectrum. 

Since atoms tend to sample larger regions of an anharmonic potential energy landscape as the lattice temperature increases, a mode dependent shift of phonon frequencies obtained in the static limit $(T=0\,\rm{K})$ occurs at finite temperatures. Corrective frequency shifts can be determined perturbatively~\cite{MF62} or, alternatively, by obtaining effective force constants at finite temperature~\cite{HAS11} which in turn enables the calculation of temperature dependent phonon frequencies and eigenvectors. As input to our BTE calculations as well as to the computation of normal mode coordinates within the EMD framework, here we consider force constants exclusively in the static limit. While some information about the significance of anharmonic corrections can be inferred from NMA derived phonon frequencies at finite temperature, a thorough investigation of the temperature dependence of phonon frequencies and eigenvectors with possible effects on TC predictions is left for future work.

\begin{figure*}
    \centering
    \subfloat[]{\label{fig:P1:a}}
	\subfloat[]{\label{fig:P1:b}}
	\subfloat[]{\label{fig:P1:c}}
	\vspace{-2\baselineskip}
    \includegraphics[width=1\textwidth]{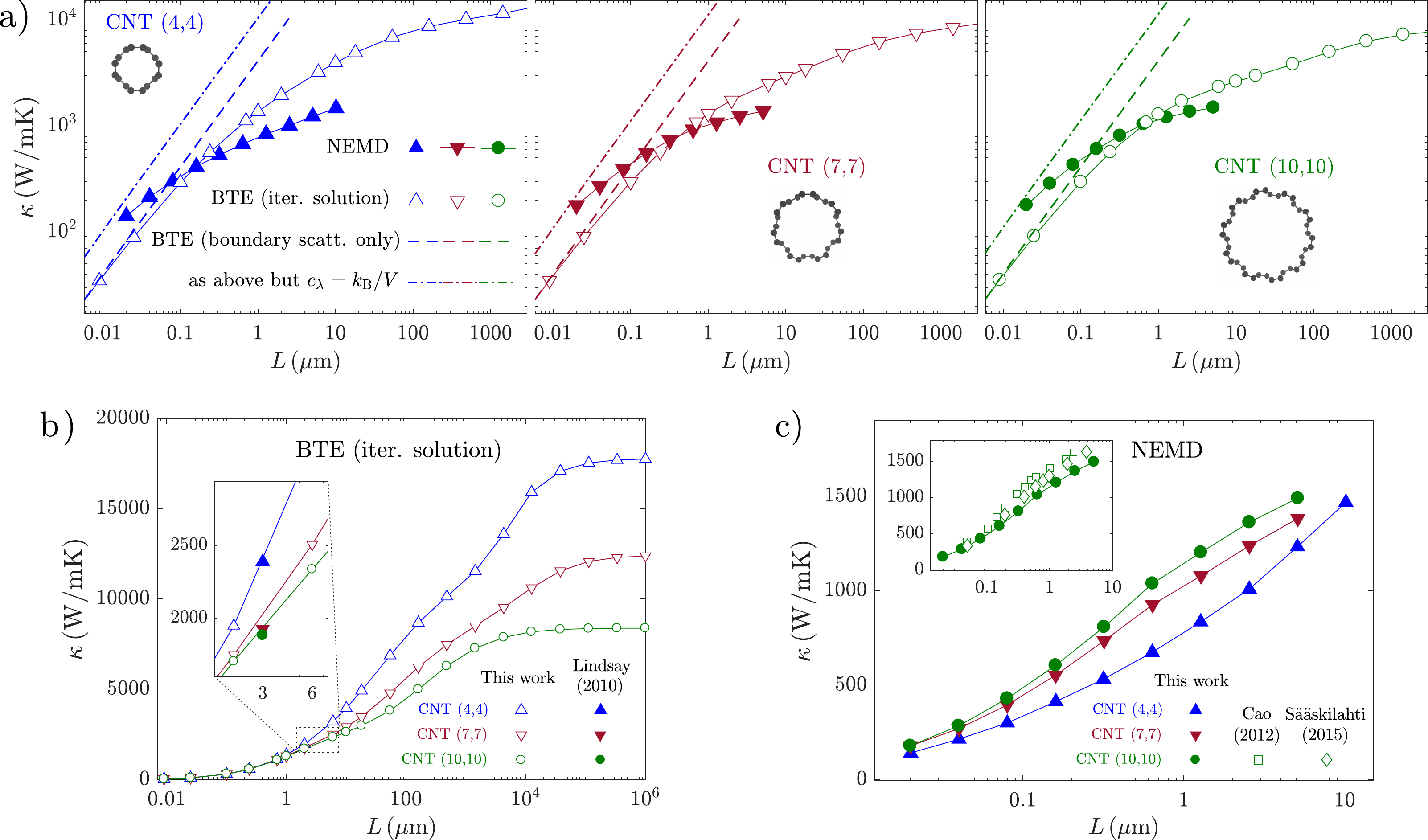}
    \caption{\label{fig:P1} Length dependence of TC as a function of tube diameter at $T=300\,\rm{K}$ arising from the BTE and NEMD approach. BTE results of this work are obtained with $N=400$ $q$-grid points. (a) Comparison of computational approaches with CNT diameter increasing from the left to the right plot. Straight lines indicate TC values that result exclusively from phonon-boundary scattering as per the BTE framework. Dashed and dashed-dotted lines highlight the role of quantum and classical heat capacity, respectively. (b) BTE results in log-linear scale with reference data taken from Lindsay et al.~\cite{LBM10}. (c) NEMD results in log-linear scale with reference data from Cao et. al.~\cite{CQ12} and from S\"a\"askilahti et al.~\cite{SOV15}.}
\end{figure*}

\section{Results and Discussion}\label{sec:RSLTS_DIS}

\subsection{Ballistic-to-diffusive transport transition from the BTE and NEMD}

In Fig.~\ref{fig:P1}, we show the TC of CNTs as predicted by the BTE and by the NEMD approach. Framework specific trends of TC with respect to the tube length are compared with each other for CNTs of different diameters in Fig.~\subref*{fig:P1:a}. Inspecting the same TC values obtained by each computational approach separately, we present the BTE and NEMD derived tube length dependence of TC in Figs.~\subref*{fig:P1:b} and~\subref*{fig:P1:c}, respectively.

For tubes of varying diameter, both calculations follow a similar trend in the ballistic regime at the submicron scale, as can be seen in Fig.~\subref*{fig:P1:a}. At this length scale, BTE predictions are only weakly affected by intrinsic three-phonon scattering and fall below corresponding TC values from NEMD. In classical MD, equipartition under the harmonic lattice approximation implies a per-phonon heat capacity of $c_{\lambda}=k_{\rm{B}}/V$ which overestimates the true (quantum) value that is enforced in the BTE framework as per Eq.~\eqref{eq:BTE}. This overprediction manifests in the ballistic transport regime, where TC from NEMD tends to be equally overpredicted.

As indicated by the dashed and dashed-dotted lines in Fig.~\subref*{fig:P1:a}, turning off phonon-phonon interaction in the BTE framework implies $\kappa(L)=\gamma L$, where the slope depends on phonon dispersion branches and on the choice of classical or quantum phonon heat capacity as $\gamma =\sum_{\lambda}c_{\lambda}|v_{\lambda}|/2$. Focussing on the tube diameter dependence of $\gamma$, we find that the use of classical heat capacity leads to ${\gamma_{(4,4)}/\gamma_{(10,10)}\approx 0.91}$, whereas the ratio increases to $\gamma_{(4,4)}/\gamma_{(10,10)}\approx 0.99$ if the quantum heat capacity is assumed. To a certain degree, these numbers might explain the fact that NEMD simulations in the ballistic regime provide larger TC values for larger diameter tubes, see Fig.~\subref*{fig:P1:c}, which is not noticeable according to BTE derived (quantum) predictions in Fig.~\subref*{fig:P1:b}.

In the limit of long tubes, anharmonic phonon-phonon scattering becomes increasingly important. Figure~\subref*{fig:P1:b} shows that BTE derived TC values first rise approximately logarithmically with tube length but converge eventually towards a finite value. Hence, an intrinsic value of TC and a diffusive transport regime are suggested by an iterative solution of the BTE under the inclusion of three-phonon scattering processes for all tube diameters considered. As we will discuss later, even though the convergence with $L$ is straightforwardly addressed in the BTE framework by setting ${(\tau^{0}_{\rm{bs},\lambda})^{-1}=2|v_{\lambda}|/L= 0}$, it remains computationally challenging in the BTE approach to pinpoint limiting TC values for $L\rightarrow \infty$ which prove to be vastly dominated by the contribution of long wavelength acoustic phonons. In NEMD, the maximally accessible system size is restricted by the atom count and by an increase of simulation time necessary to drive longer CNTs into a nonequilibrium steady-state. As can be seen in Figs.~\subref*{fig:P1:a} and \subref*{fig:P1:c}, the steady increase of TC predictions in the range up to $L=10\,\rm{\mu m}$ prevents one from making definite statements as to the existence of an intrinsic value of TC in the limit of long tubes. Focussing on the tube length dependence as a function of tube diameter in Fig.~\subref*{fig:P1:c}, a power law divergence in the case of the (4,4) tube and a weaker, possibly logarithmic, divergence of the (7,7) and (10,10) tubes might be deduced. Equally likely, NEMD results could be described by a formula of the form $\kappa(L)=\kappa^{\infty}(1+\lambda_{\rm{eff}}/L)^{-1}$ which, in accordance with BTE results, would imply a length saturation of TC with limiting value $\kappa^{\infty}$ and finite (effective) phonon mean free path $\lambda_{\rm{eff}}$.

Comparing TC predictions of both frameworks in Fig.~\subref*{fig:P1:a}, NEMD results are surpassed by BTE predictions for $L\gtrsim 1\,\rm{\mu m}$ irrespective of tube diameter. This crossover relates to the distinct treatment of resistive phonon-phonon scattering in either framework. The BTE approach truncates higher order resistive phonon-phonon scattering, whereas the NEMD approach does not account for the Bose-Einstein statistics of phonons entering into the calculation of anharmonic scattering rates as per Eq.~\eqref{eq:3rd_ampltudes}. To scrutinize the role of phonon statistics in thermal transport calculations, some studies~\cite{TLM09,HSD12,PXC19} employed classical statistics in the BTE framework by adopting phonon occupation numbers of the form $n_{\lambda}=k_{\rm{B}}T/\hbar \omega_{\lambda}$.  The high Debye temperature $(\sim 2200\,\rm{K})$ of CNTs, however, makes such an approach problematic as it was shown by Feng et al.~\cite{FR16} that anharmonic scattering rates as stated in Eqs.~\eqref{eq:LTIMES_3rd_RTA}-\eqref{eq:3rd_matrix} are well-defined only under the assumption of Bose-Einstein statistics.

\subsection{Comparison with earlier BTE results}

Previously, the BTE framework under the computation of the full spectrum of anharmonic three-phonon scattering was applied to CNTs by Lindsay et al.~\cite{LBM09,LBM10} and, more recently, by Yue et al.~\cite{YOH15}. The former authors investigated the length~\cite{LBM09} and diameter dependence~\cite{LBM10} of TC based on force constants derived from different parametrizations of the Tersoff potential. While not taking into account phonon-boundary scattering, Yue et al.~\cite{YOH15} studied the diameter dependence of TC in the diffusive regime using force constants from \textsl{ab-initio} calculations. 

As indicated by the limiting behavior of the ballistic-to-diffusive transition of heat transport in Fig.~\subref*{fig:P1:b}, even with only three-phonon scattering included, our BTE results imply a saturation of TC in tubes with length beyond the millimeter scale. A similar regime of convergence was determined by Lindsay et al.\ in~\cite{LBM09} whose diameter dependent predictions for $L=3\,\rm{\mu m}$ in~\cite{LBM10} based on the optimized Tersoff potential fall into line with our results, as highlighted by the inset of Fig.~\subref*{fig:P1:b}. In contrast to the calculations in~\cite{LBM09}, however, we do not encounter a divergence of TC by solving Eq.~\eqref{eq:iterBTE} iteratively in the diffusive regime, where $L\rightarrow \infty$, which would motivate an approximative inclusion of higher order phonon-phonon interaction as suggested by the earlier work of Mingo et al.~\cite{MB05}. Likewise, no such divergence stemming exclusively from anharmonic three-phonon scattering in the diffusive regime was reported by Yue et al.~\cite{YOH15}.

\subsection{Comparison with earlier NEMD results}

Earlier NEMD simulations that also adopted the optimized Tersoff potential to study the length dependence of TC for CNTs of varying chiralities and for a (10,10) tube, were conducted by Cao et al.~\cite{CQ12} and by S\"a\"askilahti et al.~\cite{SOV15}, respectively. The former work considered tubes in the range $L\leq 2.4\,\rm{\mu m}$ up to which TC was found to increase with tube diameter.  In the diffusive transport regime, Cao et al.\ proposed a power law divergence of TC, $\kappa=L^{\alpha}$, where the exponent $\alpha\sim 0.21$ is independent of tube diameter. As shown in Fig.~\subref*{fig:P1:c}, we similarly detect that TC increases with diameter in the range $L\lesssim 5\,\rm{\mu m}$, but that this trend might no longer hold in larger simulations due to a more pronounced increase of TC with length as observed for the (4,4) tube in comparison to the larger diameter tubes of chirality (7,7) and (10,10). The NEMD simulations of S\"a\"askilahti et al.~\cite{SOV15} focussed on a (10,10) tube with $L\leq 4\,\rm{\mu m}$. As indicated by the inset of Fig.~\subref*{fig:P1:c}, our NEMD predictions for the (10,10) tube compare satisfactorily with trends reported by the aforementioned studies. Given the prohibitive computational demands of longer tube NEMD simulations, S\"a\"askilahti et al.\ extrapolated the length dependence of TC based on a spectrally resolved phonon mean free path obtained from NEMD simulations. Here, the authors identified the importance of low frequency phonons whose dissipation rates in the limit $\omega \rightarrow 0$ were found to determine whether TC saturates or diverges with length. On a similar note, we focus on low frequency acoustic modes at the level of individual phonon branches in the following.

\subsection{The role of long wavelength acoustic phonons in the BTE framework}\label{sec:BTE_ac_phn}

\begin{figure}
\centering
\includegraphics[width=0.95\columnwidth]{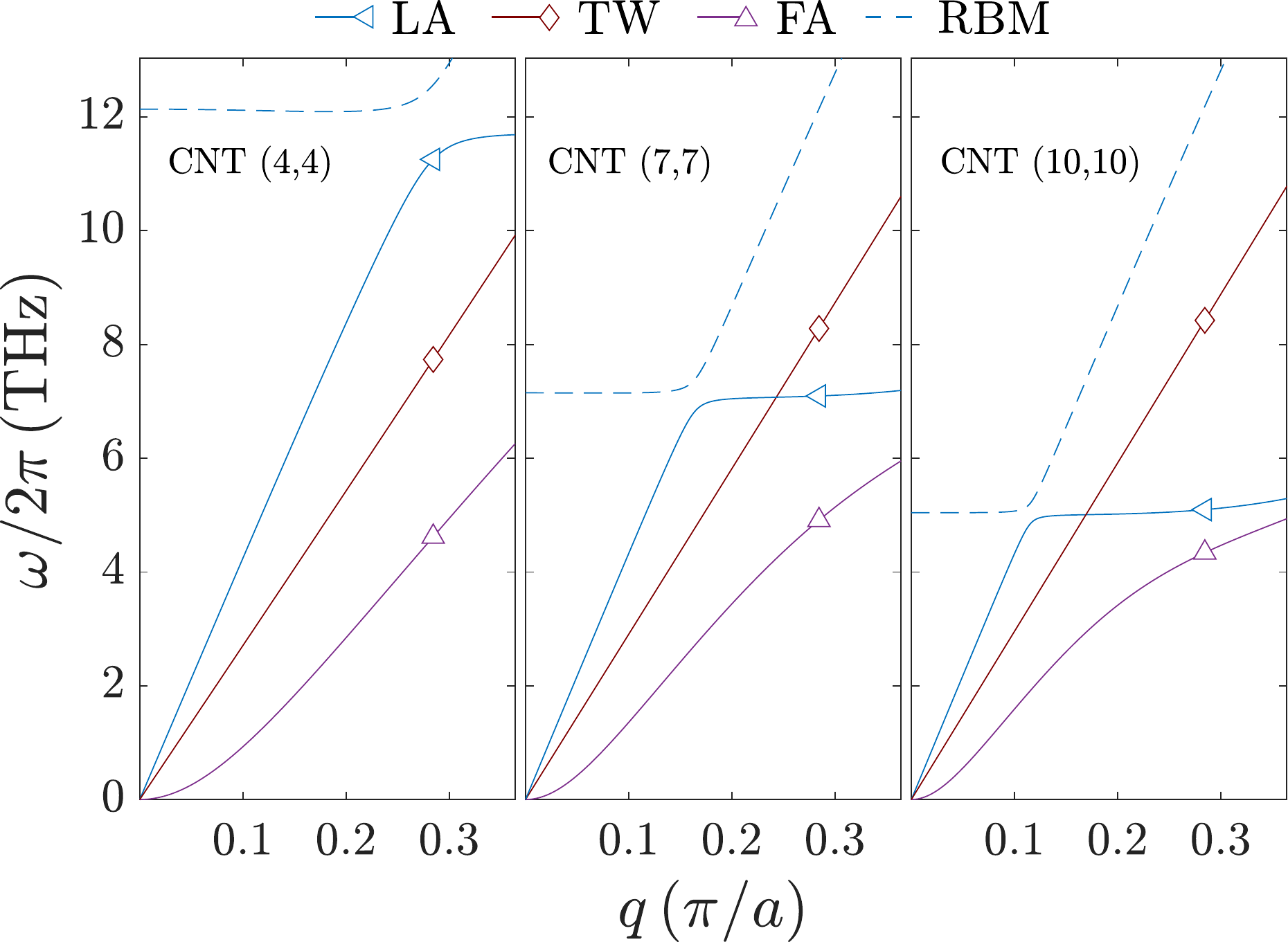}
\caption{\label{fig:P2} Acoustic phonon branches in the low frequency region for armchair CNTs of varying diameter. The longitudinal (LA) and twisting mode (TW) exhibit a linear dispersion, $\omega \sim q$, in the limit $q\rightarrow 0$, whereas the two degenerate flexural (FA) modes display $\omega \sim q^2$. Also shown is the radial breathing mode (RBM). LA and RBM modes possess equivalent symmetry which causes a mode hybridization and an avoided crossing of phonon branches.} 
\end{figure}


Motivated by the Klemens approximation of spectrally resolved three-phonon scattering rates implying ${\tau^{-1}(\omega)\sim \omega^{2}}$~\cite{K69,K58}, previous studies~\cite{MB05,CB04,WW06} ascribed a diverging trend of TC with tube length to low frequency acoustic phonons with low intrinsic attenuation. CNTs of any chirality give rise to four distinct acoustic modes. For the CNTs considered in Fig.~\ref{fig:P1}, the dispersion of acoustic modes in the low frequency region is shown in Fig.~\ref{fig:P2} as determined by a harmonic lattice calculation. The longitudinal (LA) and twisting (TW) modes exhibit a linear dispersion, $\omega\sim q$, whereas the two degenerate flexural modes follow a quadratic dependence, $\omega \sim q^2$, for small wavenumbers. 

Within the BTE framework, the per-phonon TC contribution of acoustic phonons in the low frequency limit, $\hbar \omega/k_{\rm{B}}T \rightarrow 0$, becomes ${\kappa_{\lambda}=1/V(k_{\rm{B}}v_{x,\lambda}^2\tau_{\lambda}})$. Considering the continuum limit of Eq.~\eqref{eq:BTE}, an integral over wavenumbers determines the per-branch TC contribution of acoustic phonons, ${\kappa_j\sim \int_0 v_{x,\lambda}^{2} \tau_{\lambda}}{\rm{d}}q$. 
For the linear LA and TW branches, the phonon group velocity as given by the slope of each branch in the limit $q\rightarrow 0$ yields a constant,  $v_{x,\lambda}=c$, so that the scaling of TC can be estimated as
\begin{align}
\kappa_{\rm{LA}/\rm{TW}}(L)\sim\int_0\frac{1}{q^a+c/L}{\rm{d}}q \sim \begin{cases}
 L^{1-1/a}, & a\neq1\\
\ln L, & a=1
\end{cases}
\end{align}
in the limit $L\rightarrow \infty$, where a power law of inverse anharmonic lifetimes, $\tau_{\lambda}^{-1}\sim ~ q^{a}$, in the presence of phonon-boundary scattering, $\tau^{-1}_{\rm{bs},\lambda}\sim v_{x,\lambda}/L$, is assumed. A similar argument applies to the TC contribution of FA modes whose group velocity scales linearly with wavenumber, $v_{x,\lambda}\sim q$, so that
\begin{align}
\kappa_{\rm{FA}}(L)\sim\int_0\frac{q^2}{q^a+q/L}{\rm{d}}q \sim \begin{cases}
 L^{1-2/(a-1)}, & a\neq3\\
 \ln L, & a=3
\end{cases}
\end{align}
in the limit $L\rightarrow \infty$. Consequently, for CNTs, a divergence of TC with tube length can be demonstrated by detecting a power law divergence of acoustic phonon lifetimes, $\tau_{\lambda}\sim q^{-a}$, with exponent $a\geq 1$ or $a\geq 3$ in case of linear or quadratic phonon dispersion, respectively.

\begin{figure}
\centering
\includegraphics[width=0.98\columnwidth]{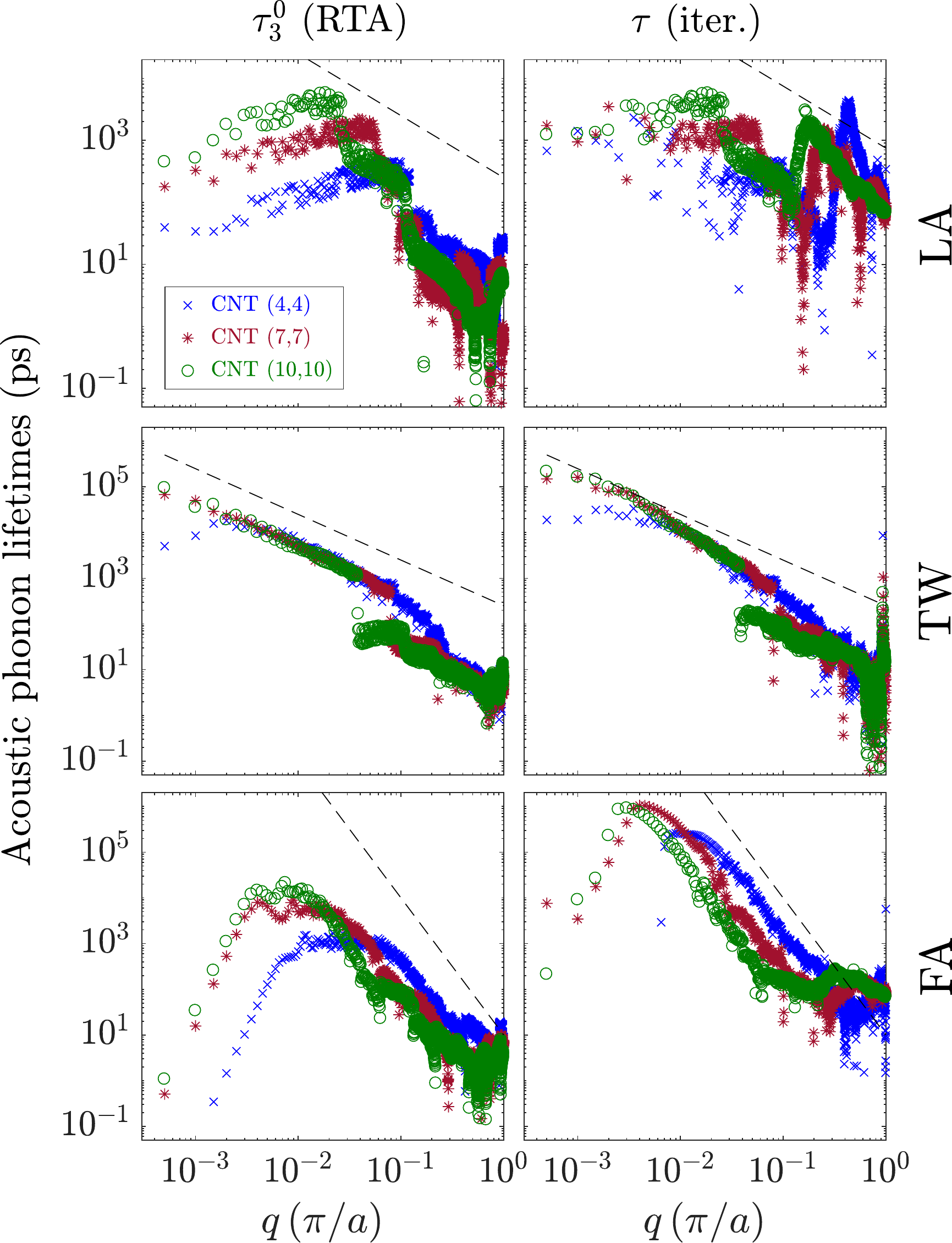}
\caption{\label{fig:P3} Lifetimes of acoustic modes at $T=300\,\rm{K}$ as a function of wavenumber arising from anharmonic three-phonon scattering in the BTE framework for tubes of varying diameter. Lifetimes from an RTA and from an iterative solution of the BTE are shown in the first and second column, respectively. On the log-log scale, the critical slope of a power law $\tau\sim q^{-a}$ inducing a divergent TC contribution in the limit $L\rightarrow \infty$ is indicated by the dashed lines.} 
\end{figure} 

Resulting from the full spectrum of anharmonic three-phonon scattering in CNTs, we show BTE derived intrinsic lifetimes of acoustic phonons both under RTA and stemming from an iterative solution of Eq.~\eqref{eq:iterBTE} in Fig.~\ref{fig:P3}. Applying an RTA, no conclusive evidence of a power law causing a divergence of TC can be found. Decreasing the wavenumber of LA and FA modes, lifetimes saturate and a decreasing trend becomes apparent for sufficiently small wavenumbers. This turn-around manifests at a wavenumber that decreases with increasing tube diameter. In comparison to LA and FA mode lifetimes under RTA, TW mode lifetimes exhibit a monotonic rise down to significantly smaller wavenumbers. Here, lifetimes of the (4,4) tube indicate a saturation, whereas TW lifetimes in tubes of larger diameter might exhibit a saturation that lies beyond the range of resolved wavenumbers. Attributed to an increase of normal three-phonon processes, finite acoustic phonon lifetimes under RTA in the limit $q\rightarrow 0$ were reported earlier for CNTs~\cite{LBM09} and are consistent with our results. Accounting for the nondissipative character of normal scattering processes, Fig.~\ref{fig:P3} equally hints at a saturation of iterated lifetimes in the limit $q\rightarrow 0$ so that a finite TC contribution of acoustic modes in the limit $L\rightarrow \infty$ is implied. It is noteworthy, however, that iterated lifetimes of FA modes rise by approximately two orders of magnitude as compared to the corresponding RTA prediction. We also find that iterated lifetimes of FA modes are prone to numerical uncertainties in the limit of small wavenumbers. Specifically, in the case of the (4,4) and (7,7) tubes, iterated lifetimes of the FA mode obtained for different $q$-number discretizations, $\Delta q=2\pi/Na$, are found to vary drastically in the range $0<q<0.01\pi/a$ depending on the chosen value of~$N$, see Fig.~S4 of the Supplemental Material~\cite{Note1}.

\subsection{Limiting TC values from the BTE and the wavenumber discretization error}

\begin{figure}
\centering
\includegraphics[width=0.95\columnwidth]{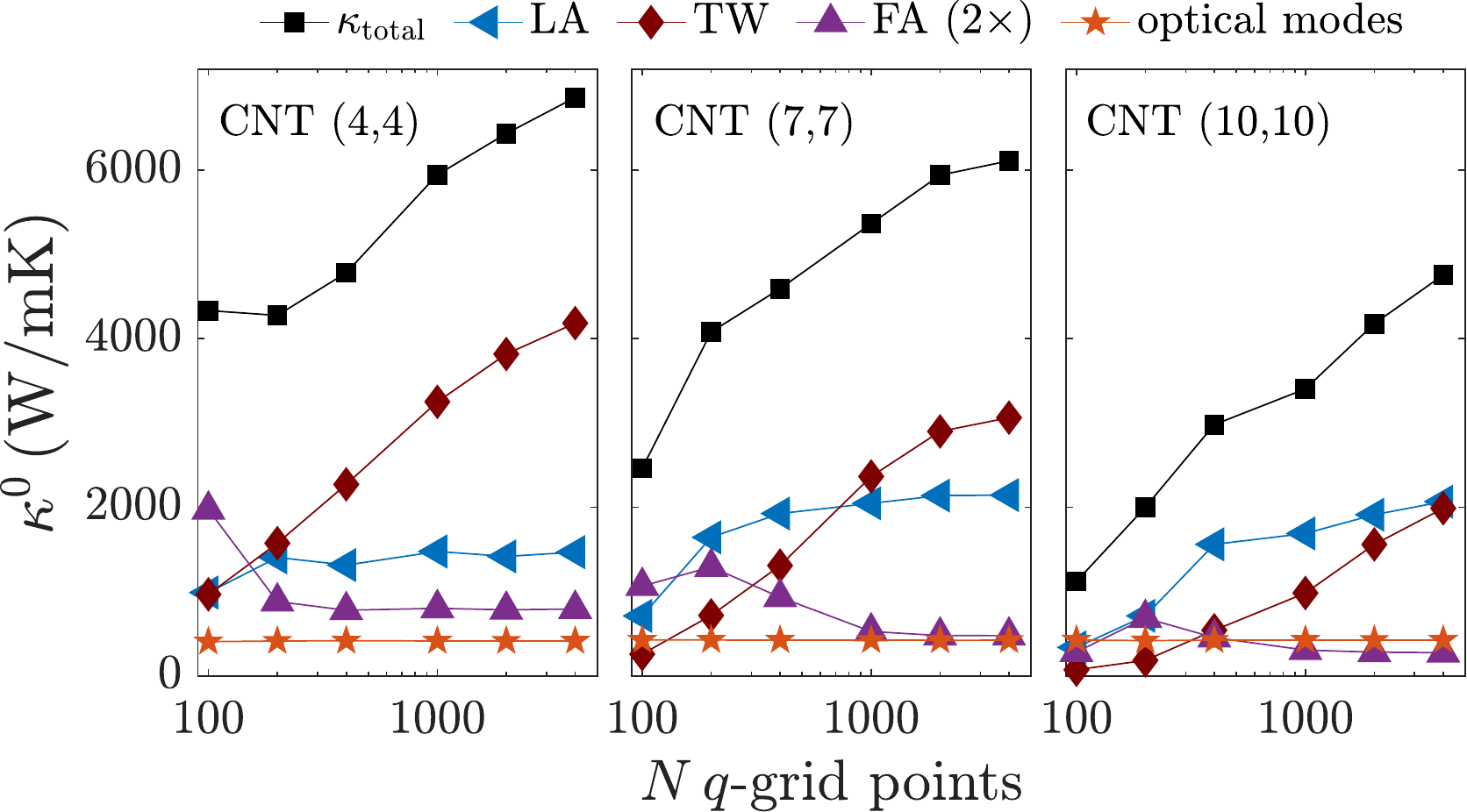}
\caption{\label{fig:P4} BTE predictions of phonon branch-resolved TC under RTA at $T=300\,\rm{K}$ as a function of tube diameter and wavenumber grid resolution. Considering the long tube limit, $L\rightarrow \infty$, the plot shows finite-size effects that result from a truncation of long wavelength modes with ${q<2\pi/Na}$.} 
\end{figure}

With anharmonic phonon lifetimes in hand, it is instructive to examine the individual mode-by-mode contributions to TC in the long tube limit by setting $(\tau^0_{{\rm{bs}},\lambda})^{-1}=0$. In Fig.~\ref{fig:P4}, we show per-branch TC contributions under RTA, ${\kappa_j^0=\sum_{q}c_{\lambda}v_{x,\lambda}^2\tau^0_{3,\lambda}}$, with respect to the employed wavenumber discretization. Since long wavelength modes with $0<q<2\pi/Na$ are truncated by an equidistant discretization of wavenumbers, $\Delta q=2\pi/Na$, sufficiently large values of $N$ are required to capture the dominant TC contribution stemming from long wavelength acoustic phonons. Given the relatively large number of phonon branches in CNTs, however, and a correspondingly large number of three-phonon scattering processes, we find that BTE calculations beyond $N>4000$ become computationally impracticable. Hence, as shown in Fig.~\ref{fig:P4}, sufficiently dense $q$-grids to achieve a saturated TC contribution from TW modes are in fact not obtained. Even though a tube length saturation of TC is evidenced by the scaling of acoustic phonon lifetimes, which are resolved down to the smallest computable wavenumber in Fig.~\ref{fig:P3}, reported values of TC in the diffusive limit, $L\rightarrow\infty$, remain subject to wavenumber discretization errors. Moreover, we show in Sec.~II of the Supplemental Material~\cite{Note1} that TC predictions in the diffusive limit derived from an iterative solution of the BTE are not strictly increasing with $q$-grid discretization factor $N$, which may hinder an extrapolation to the wavenumber continuum limit. Despite the uncertainties related to TC predictions in the long tube limit, we expect the general trends of TC with respect to both tube length and diameter of CNTs reported in Fig.~\subref*{fig:P1:b} to be robust.


\subsection{NMA of long wavelength acoustic phonons}\label{sec:NMA_OF_ACS}

\begin{figure*}
\centering
\includegraphics[width=2\columnwidth]{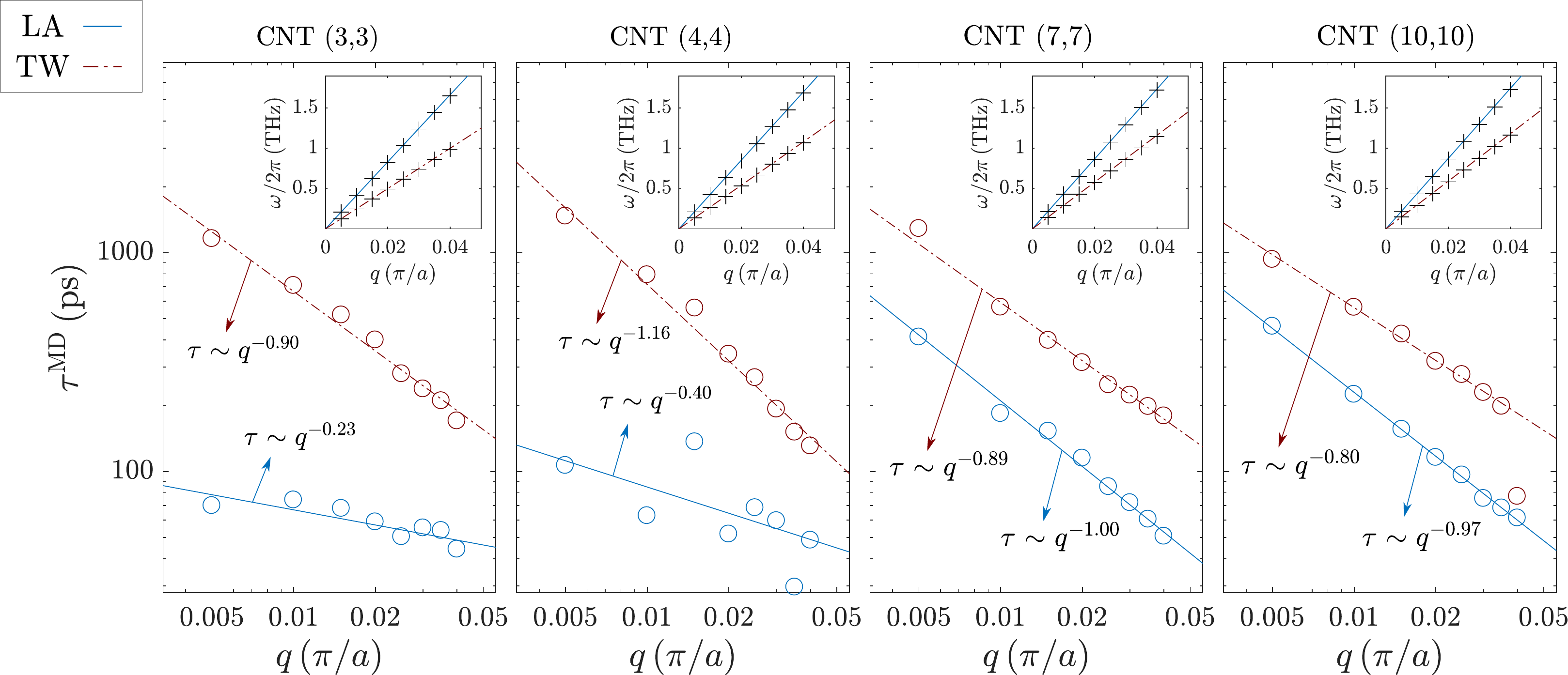}
\caption{\label{fig:P5} Phonon lifetimes of  longitudinal (LA) and twisting (TW) modes resulting from normal mode analysis (NMA) in the limit of small wavenumbers. Plots are arranged in order of increasing diameter (from left to right). Underlying EMD simulations at $T=300\,\rm{K}$ comprise $N=400$ translational unit cells. Insets display the corresponding linear phonon dispersion with NMA derived finite temperature frequencies shown by plus markers.} 
\end{figure*}

Given BTE derived trends of acoustic phonon lifetimes, NMA in the framework of EMD allows to probe whether these trends persist if higher order phonon-phonon interaction is present. In analogy to BTE calculations, MD simulations of CNTs with $N$ translational unit cells under periodic boundary conditions equally imply a discretization of wavenumbers ${\Delta q=2\pi/Na}$ and a truncation of phonon modes with wavelength $\Lambda > Na$. With the objective of detecting acoustic phonons in a regime of desirably long wavelength by means of NMA, we conduct EMD simulations of tubes with $N=400$ unit cells. By further restricting the on-the-fly calculation of acoustic normal mode coordinates during EMD runtime to the eight smallest wavenumbers, $q=(2\pi/400a)\,m$ with $m=(1,\dots,8)$, we achieve an acceptable trade-off between system size and simulation runtime as detailed in Sec.~\ref{sec:CMPT_DTLS}. Here, it is also worth mentioning that the LA mode can only be identified as such in simulations of sufficiently long tubes since a mode hybridization with the radial breathing mode occurs at a diameter dependent wavenumber, see Fig.~\ref{fig:P2}. For instance, enabling the LA mode of the (10,10) tube with $N_{\rm{LA}}$ distinct wavenumbers requires the simulation of at least $N=20\times N_{\rm{LA}}$ translational unit cells.

 We provide our NMA raw data in the form of mode resolved frequency spectra including fit lines as per Eq.~(\ref{eq:KIN_E}) in Sec.~III of the Supplemental Material~\cite{Note1}. Remarkably, we find that frequency spectra corresponding to FA modes of all considered diameters display no distinctive Lorentzian line profile at the lower end of wavenumbers, where harmonic mode frequencies fall below $\omega_{\lambda}/2\pi<0.3\,\rm{THz}$. Supporting BTE derived trends of FA phonon lifetimes in the limit $q\rightarrow 0$, we attribute an absence of well-defined frequency peaks to an increase of mode-specific dissipation rates stemming from normal and umklapp scattering.
 

For acoustic modes belonging to the LA and TW branches, NMA predictions of phonon lifetimes are shown in Fig.~\ref{fig:P5}. Here, we observe that the LA mode exhibits stronger attenuation in tubes of chirality (3,3) and (4,4) as compared to the larger diameter (7,7) and (10,10) tubes. This observation falls into line with BTE derived LA lifetimes in Fig.~\ref{fig:P3} which are found to reach larger values in tubes of larger diameter. Fit lines $\tau_{\rm{LA}}\sim q^{-b}$ with significantly smaller exponent $b$ in lower diameter tubes might be ascribed to an onset of a saturation in the limit $q\rightarrow 0$ which occurs for larger values of $q$ in smaller diameter tubes as predicted by the BTE framework. In the case of the TW branch, we detect a monotonic increase of phonon lifetimes down to the smallest resolvable wavenumber irrespective of tube diameter. Fit lines ${\tau_{\rm{TW}}\sim q^{-b}}$ indicate that exponents lie in the range from $0.80$ to $1.16$, which conforms with an earlier NMA derived prediction of ${\tau_{\rm{TW}}\sim q^{-1.1}}$  based on a (10,10) tube~\cite{OPS11}. Given the spread of measured exponents in Fig.~\ref{fig:P5}, we refrain from relating the power law exponent of NMA derived TW lifetimes with a possibly diameter dependent power law divergence of TC as observed within NEMD simulations. Such an attempt would be further obstructed by the fact that normal (nondissipative) and umklapp (dissipative) scattering are treated on an equal footing so that scattering rates from NMA tend to overestimate the true energy dissipation rates which are of relevance for heat conduction in NEMD. 

\begin{figure}
\centering
\includegraphics[width=0.88\columnwidth]{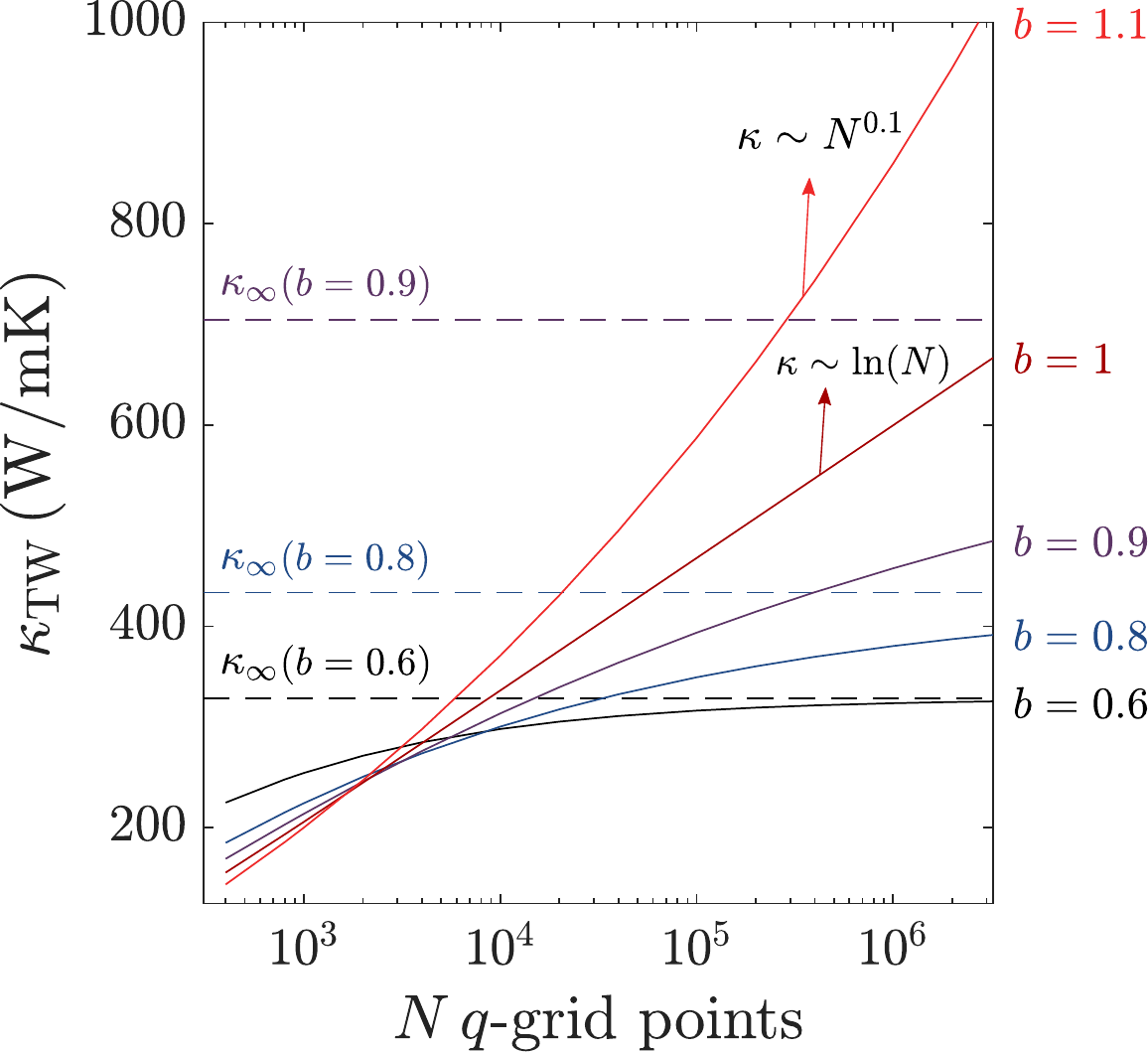}
\caption{\label{fig:P6} Thermal conductivity contribution of the TW mode in the EMD framework based on a power law extrapolation of TW phonon lifetimes, $\tau=\beta q^{-b}$. The number of $q$-grid points $N$, as shown on the horizontal axis, is proportional to the length of the simulated tube in EMD. Absolute TC values are determined by adopting the cross sectional area of the (7,7) tube and by further assuming that $\tau = 1\,\rm{ns}$ at $q=0.005 \,\pi/a$.}
\end{figure}

\subsection{The role of long wavelength acoustic phonons in the EMD framework}

Considering predictions from NMA as a lower bound to TC in the absence of phonon-boundary scattering, the consequences of an observed power law, ${\tau_{\rm{TW}}\sim q^{-b}}$, extending beyond the picosecond range are profound. 
Taking the (7,7) tube as an example and assuming that a power law of TW mode lifetimes remains intact in tubes with $N>400$ unit cells, Fig.~\ref{fig:P6} illustrates the hypothetical tube length dependence of TC predictions in the framework of EMD resulting from TW modes with ever increasing wavelength. Notably, even if lifetimes continue to follow a strict power law with exponent $b<1$, the simulation of sufficiently long tubes reaching converged TC values would quickly become impractical. In fact, since there is no additional attenuation of phonon lifetimes caused by phonon-boundary scattering in the EMD framework, a power law divergence of acoustic phonon lifetimes would lead to an even more pronounced tube length dependence of TC predictions in the EMD as compared to the NEMD framework.

 In light of $q$-grid saturated lifetimes in the framework of the BTE, one might conclude that it is a breakdown of the power law dependence of lifetimes, $\tau_{\lambda}\sim q^{-b}$, in the limit $q\rightarrow 0$ which lies at the heart of the tube length convergence of phononic TC in CNTs. Such a breakdown would be consistent with other calculations of TC in the framework of EMD which did not find a significant variation of TC with simulated tube length \cite{FDH19,FPW15}. Our NMA data indicates that a breakdown of the power law divergence of acoustic phonons occurs for quadratic FA modes in CNTs of all considered diameters. In Fig.~\ref{fig:P5}, LA modes point in this direction by signifying a weakening of the lifetime divergence in tubes of chirality (3,3) and (4,4). At the same time, however, a pronounced increase of TW phonon lifetimes down to the smallest resolvable wavenumber is confirmed by both NMA and BTE calculations. In order to capture the role of heat carrying TW modes in CNTs, Fig.~\ref{fig:P5} implies that simulated tubes in EMD should comprise at least $N\gtrsim 400$ unit cells. To avoid wavenumber truncations errors in EMD, an even higher number of simulated unit cells, $N\gtrsim 4000$, is suggested by BTE derived TW lifetimes in Fig.~\ref{fig:P3}.

\section{Summary and Conclusion}\label{sec:CNCLS}

In this article, we have addressed thermal transport in CNTs and the conundrum of a domain size dependent TC in these low-dimensional systems by comparing results from three different atomic-level approaches to lattice heat transport. Our framework-specific findings can be summarized as follows.

(1) The BTE approach under the inclusion of the full spectrum of three-phonon scattering suggests a ballistic-to-diffusive transition of phonon mediated heat transport of the form $\kappa(L)=\kappa^{\infty}(1+\lambda_{\rm{eff}}/L)^{-1}$, where the diffusive limiting value of TC, $\kappa^{\infty}$, and an effective phonon mean free path, $\lambda_{\rm{eff}}$, are found to increase with decreasing tube diameter. The convergence of TC with tube length is predicted to occur only at macroscopic scales, requiring $L \gtrsim 1\,\rm{mm}$. Even though it proves to be numerically challenging to give accurate limiting values of TC in the diffusive transport regime, the saturation of TC itself is evidenced by acoustic phonon lifetimes $\tau_{\lambda}$ whose scaling in the long wavelength limit $q\rightarrow 0$ signifies a breakdown of the commonly assumed power law $\tau_{\lambda}\sim q^{-a}$.

(2) NEMD simulations of lattice TC reveal a violation of a quantum upper bound to TC in the ballistic transport regime for CNTs of lengths $L<0.1\,\rm{\mu m}$. Extending NEMD simulations to micrometer long tubes comprising $\sim 10^6$ atoms, a steady increase of TC is observed such that no definite statement as to the existence of a diffusive transport regime can be made. Regarding NEMD results in isolation, finite length predictions of TC can be extrapolated beyond the micrometer range using both converging and diverging trends.

(3) NMA in the framework of EMD allows one to probe BTE derived trends of acoustic phonon lifetimes in the presence of higher order phonon-phonon interaction. Supporting our findings in the BTE framework, the following qualitative trends can be inferred from NMA in the long wavelength limit $q\rightarrow 0$. A breakdown of the power law divergence of acoustic FA mode lifetimes is indicated by the lack of distinctive peaks in the corresponding polarization resolved frequency spectrum. For LA modes, a flattening of lifetimes is observed for small diameter tubes. In accordance with BTE results and irrespective of tube diameter, a pronounced increase of TW mode lifetimes hints at the predominant role of TW modes in CNT heat transport. Aiming at thermal transport properties of CNTs, our NMA results further suggest that EMD simulations should comprise $N\gtrsim 400$ translational unit cells to not truncate the transport contribution of long wavelength TW modes.

In conclusion, our results clarify some of the conflicting predictions in the literature on CNT thermal transport. Moreover, our work may serve as a starting point for future research on phonon meadiated heat transfer in CNTs and in other quasi one-dimensional materials. For example, applying the BTE framework to two-dimensional graphene, the significance of four-phonon scattering at various temperatures~\cite{FR18} as well as finite temperature corrections to both harmonic and anharmonic force constants~\cite{GFB19} have already been considered but remain to be explored in the case of CNTs.

\begin{acknowledgements}
The authors gratefully acknowledge financial support from the Natural Sciences and Engineering Research Council of Canada. J.R.\ also thanks the Alexander von Humboldt Foundation for financial support. D.B.\ was supported by the QuEST scholarship at the University of British Columbia. High performance computing resources were provided by ComputeCanada and the Quantum Matter Institute at the University of British Columbia. This research was undertaken thanks, in part, to funding from the Canada First Research Excellence Fund, Quantum Materials and Future Technologies Program.
\end{acknowledgements}

\bibliography{CNTpaper}

\end{document}


\title{Supplemental Material -\\Heat transport in carbon nanotubes: Length dependence of phononic conductivity from the Boltzmann transport equation and molecular dynamics}
\author{Daniel Bruns}
\affiliation{Department of Physics and Astronomy, University of British Columbia, Vancouver, BC, Canada V6T 1Z1}
\affiliation{Quantum Matter Institute, University of British Columbia, Vancouver, BC, Canada V6T 1Z4}
\author{Alireza Nojeh}
\affiliation{Department of Electrical and Computer Engineering, University of British Columbia, Vancouver, BC, Canada V6T 1Z4}
\affiliation{Quantum Matter Institute, University of British Columbia, Vancouver, BC, Canada V6T 1Z4}
\author{A. Srikantha Phani}
\affiliation{Department of Mechanical Engineering, University of British Columbia, Vancouver, BC, Canada V6T 1Z4}
\author{J\"org Rottler}
\affiliation{Department of Physics and Astronomy, University of British Columbia, Vancouver, BC, Canada V6T 1Z1}
\affiliation{Quantum Matter Institute, University of British Columbia, Vancouver, BC, Canada V6T 1Z4}

\maketitle


\section{NEMD: Setup and steady state temperature profiles}

\begin{figure*}[h!]
\centering
\includegraphics[width=0.7\columnwidth]{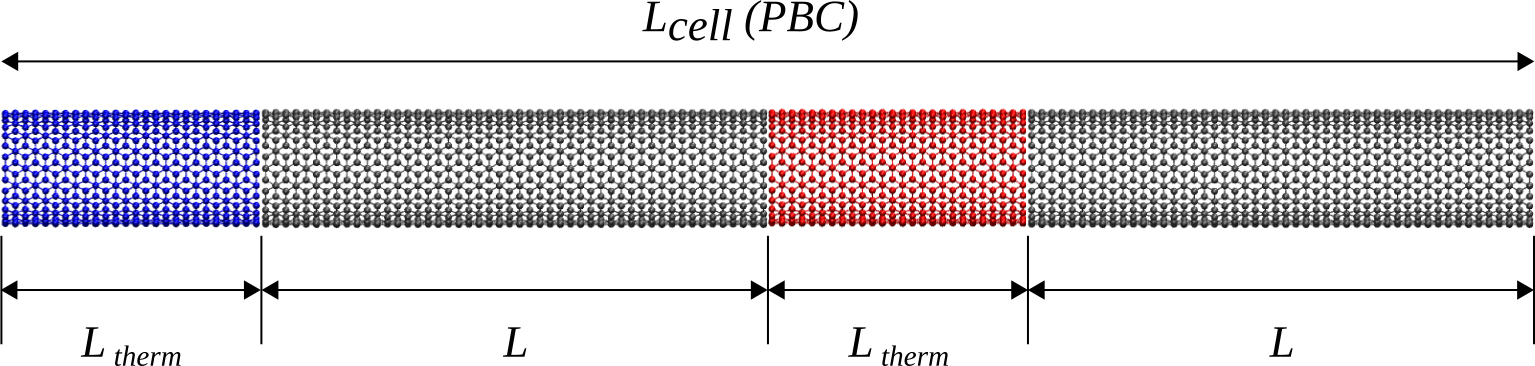}
\caption{\label{fig:P1} Simulation setup in the NEMD method. Atomic velocities are continuously swapped between the cold (blue) and hot (red) region. Under periodic boundary conditions (PBC), a triangular shaped temperature profile is induced.}
\end{figure*}

A schematic illustration of the NEMD setup is given in Fig.~\ref{fig:P1}. Applying periodic boundary conditions along the heat transport direction, the tube length $L$ as reported in the main article is given by $L=L_{\rm{cell}}/2-L_{\rm{therm}}$, where $L_{\rm{cell}}$ denotes the total length of the simulation cell along the tube axis and $L_{\rm{therm}}$ measures the length of the slabs whose atoms are subject to velocity swapping. Throughout our simulations, we impose $L_{\rm{therm}}=20\,\rm{nm}$ which corresponds to 80 translational unit cells. 

\begin{figure*}[hb]
\centering
\includegraphics[width=1\columnwidth]{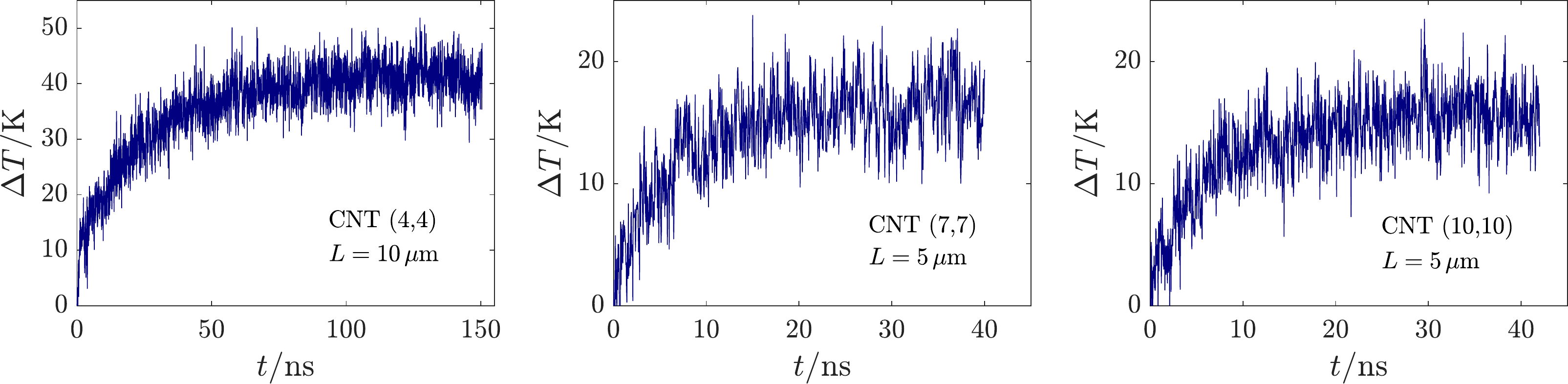}
\caption{\label{fig:P2} Temperature difference $\Delta T$ between the hot and cold slabs of the longest simulated CNTs during the steady state equilibration stage.}
\end{figure*}

To reach a regime of steady state heat transport, preceding the data production stage, an equilibration stage with a constant frequency of velocity swaps has to be conducted whose duration typically increases with the length of the simulated tube. For our largest simulations, we exemplify the onset of a steady state heat transport regime as indicated by the temperature difference $\Delta T$ between the hot and cold regions of the tube in Fig.~\ref{fig:P2}.  

\begin{figure*}[ht]
\centering
\includegraphics[width=0.95\columnwidth]{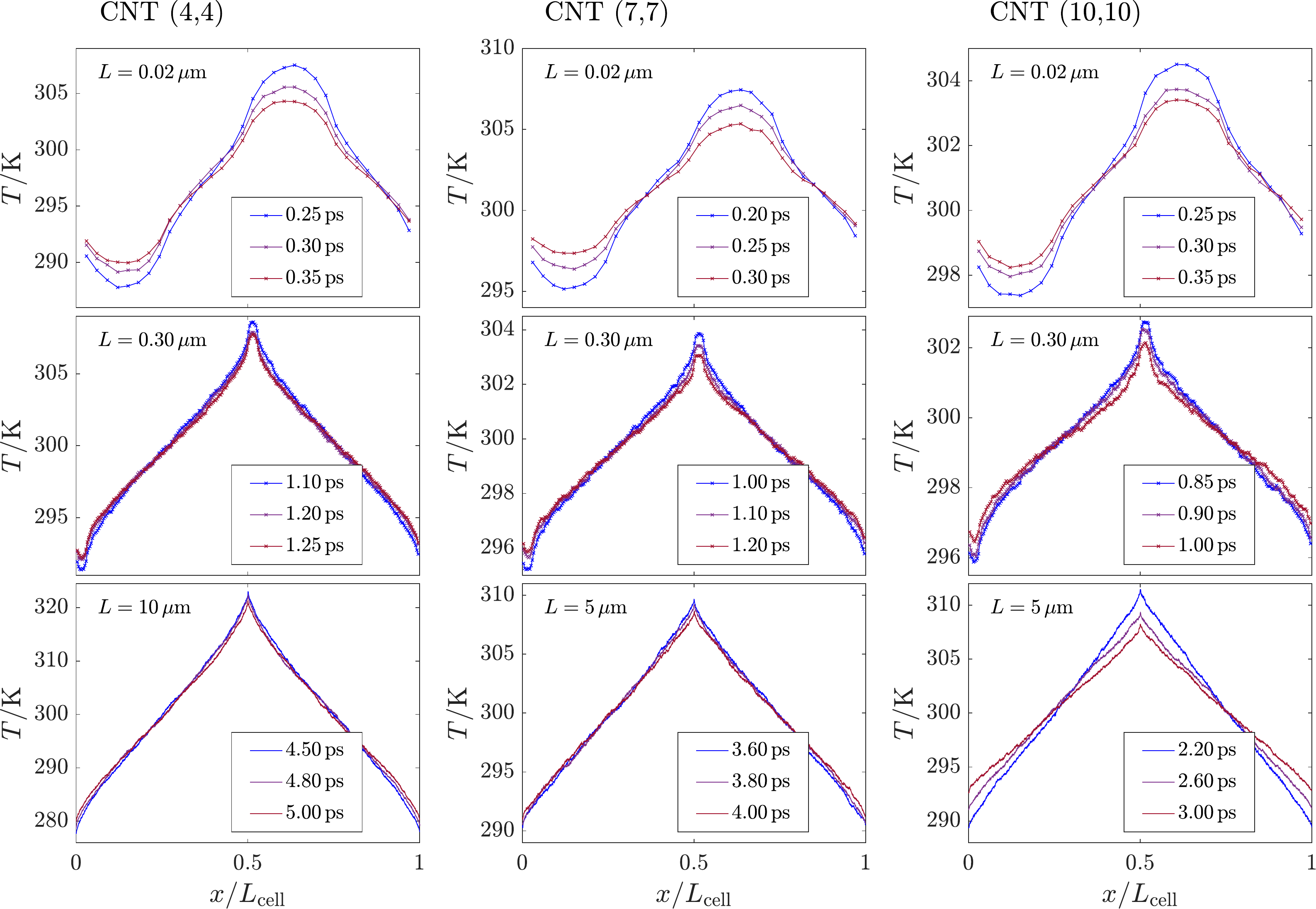}
\caption{\label{fig:P3} Time averaged steady state temperature profiles obtained within the NEMD method for CNTs of different length and diameter. For each respective configuration, three different profiles are shown that stem from independent simulations with different velocity swap periods (shown by insets).}
\end{figure*}

Temperatures are monitored in $M$ equally spaced bins along the tube, where the width of each bin is chosen to comprise ten translational unit cells. Once the steady state heat transport regime is established, snapshots of the temperature profile are taken every ${t'=10\times T_{\rm{swap}}}$, where $T_{\rm{swap}}$ stands for the imposed period of velocity swaps. Final temperature profiles are then obtained by averaging over 5000 snapshots in the case of tubes with $L\leq 1\,\rm{\mu m}$ and over 1000 snapshots in the case of tubes with $L> 1\,\rm{\mu m}$. Given final temperature profiles as shown in Fig.~\ref{fig:P3}, the temperature within the cold and hot regions are taken as average values over respective bin temperatures $T(x_{1}),\dots,T(x_{8})$ and $T(x_{M/2+1}),\dots,T(x_{M/2+8})$, respectively.

Three independent estimates of thermal conductivity (TC) are obtained for each tube diameter and length by performing three independent simulations with slightly varied velocity swap periods $T_{\rm{swap}}$. Here, we find that derived TC values do not exhibit a systematic variation with velocity swap period, suggesting that heat transport is in the linear regime. Considering the variation of TC values, the standard uncertainty of TC as given by the standard deviation divided by the square root of the number of independent runs is smaller than the plot marker size in Fig.~1 of the main article. 

Dealing with atom counts on the order of $\sim 10^6$, we use the GPU-accelerated implementation of the Tersoff potential~\cite{N17} available in the \textsl{LAMMPS} package~\cite{P95} which allows us to leverage a hybrid MPI/GPU acceleration of MD simulations on high performance compute clusters.

\section{Iterative solution of the BTE: Numerical artifacts stemming from long wavelength flexural modes}

\begin{figure}[ht]
\vspace{0.5cm}
\centering
\includegraphics[width=0.8\columnwidth]{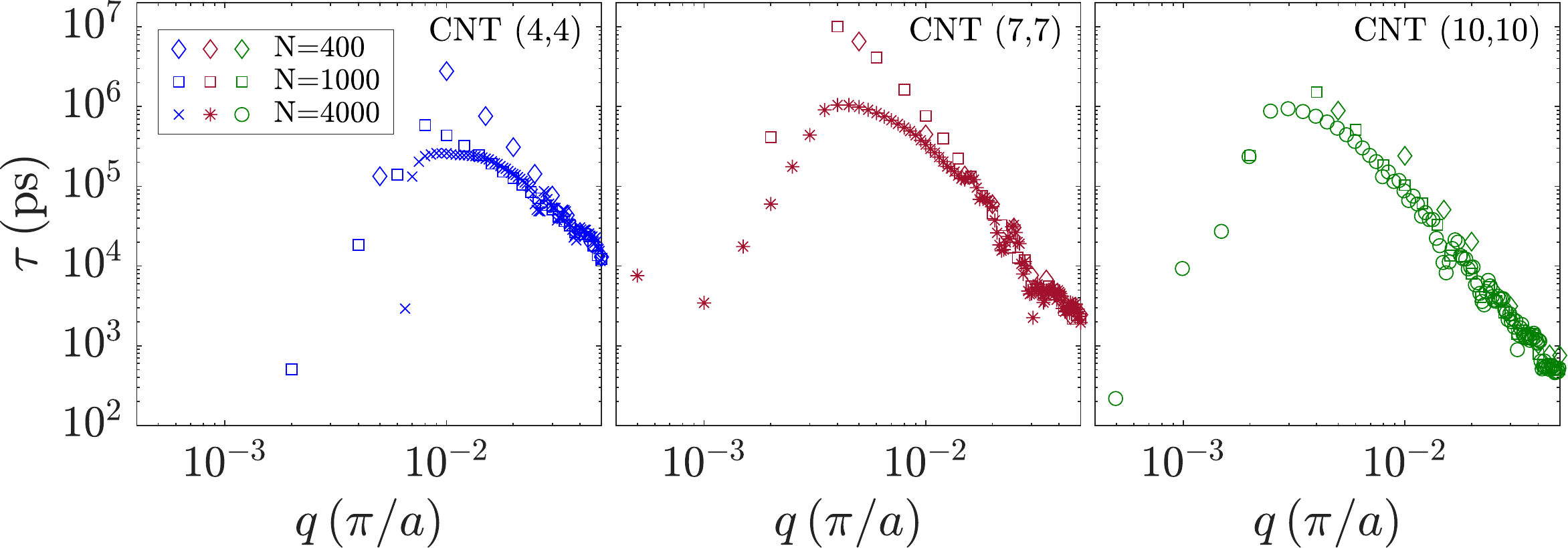}
\caption{\label{fig:P4} Iterated lifetimes of long wavelength flexural modes at $T=300\,\rm{K}$ as a function of wavenumber resulting from anharmonic three-phonon scattering. Shown are BTE results obtained for three different $q$-grid discretizations $\Delta q=2\pi/Na$, where $N\in\{400,1000,4000\}$.} 
\end{figure}

Numerical uncertainties inherent to iterated lifetimes of long wavelength flexural acoustic (FA) modes are shown in Fig.~\ref{fig:P4}. Iterated lifetimes presented in Fig.~3 of the main article result from a calculation including $N=4000$ $q$-grid points. Choosing a coarser wavenumber grid by decreasing the wavenumber count $N$ significantly alters the prediction 
of iterated FA lifetimes in the limit of small wavenumbers. This effect is more pronounced in smaller diameter tubes which is correlated with a smaller curvature of FA phonon dispersion branches in these tubes (see Fig.~2 of the main article) giving rise to an extended region of low phonon frequencies and group velocities.

\begin{figure}[ht]
\vspace{0.5cm}
\centering
\includegraphics[width=0.55\columnwidth]{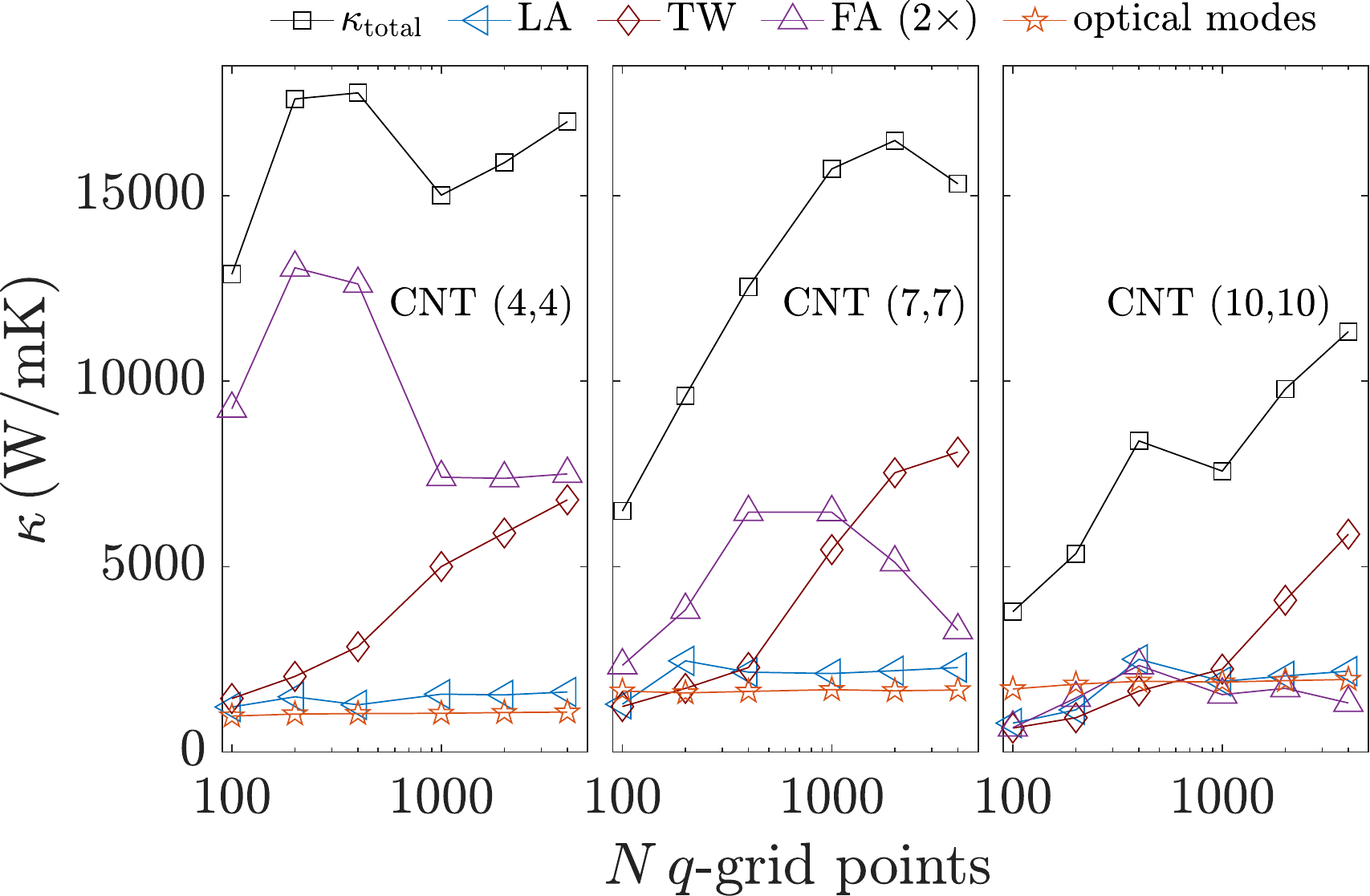}
\caption{\label{fig:P5} Phonon branch-resolved TC contributions at $T=300\,\rm{K}$ in the long tube limit derived from an iterative solution of the BTE under three-phonon scattering. The horizontal axis displays the $q$-grid discretization as given by the number $N$ of equidistant wavenumbers  $q \in (-\pi/a,\,\pi/a]$.} 
\end{figure}

Demonstrating the consequences of Fig.~\ref{fig:P4}, the per-branch contributions to TC derived from an iterative solution of the BTE for infinitely long CNTs, $L\rightarrow \infty$, are shown in Fig.~\ref{fig:P5} with respect to the employed $q$-grid discretization. Given the large spread of iterated FA mode lifetimes, no obvious convergence region of total TC with respect to the employed $q$-number discretization can be found in the case of the (4,4) and (7,7) tubes. Thus, some uncertainty of absolute TC values in the diffusive regime, $L\rightarrow \infty$, persists. 

Since CNTs of chirality $(n,n)$ give rise to $12n$ phonon branches per $q$-grid point, it should be emphasized that BTE calculations on CNTs have to account for an enormous number of allowed anharmonic three-phonon scattering processes such that finite computational resources impose an upper bound on the $q$-grid discretization factor $N$. For example, in the case of the (10,10) tube, we find a number of allowed phonon absorption processes that scales as $N_{\text{tot}}^{+}\approx 3\times 10^{5}\,N$, which leads to an overflow of counter variables in the software package \textit{shengBTE}~\cite{LCK14} if computations are carried out with $N>4000$. Going beyond this limitation, BTE calculations on CNTs can be made more resource-efficient by imposing additional selection rules on three-phonon scattering processes that derive from the rotational symmetry of CNTs, as demonstrated by the work of Lindsay et al.~\cite{LBM09}.

\section{Normal Mode Analysis of Long Wavelength Acoustic Phonons}

\begin{figure*}[h!]
\vspace{1.10cm}
\centering
\includegraphics[width=0.95\columnwidth]{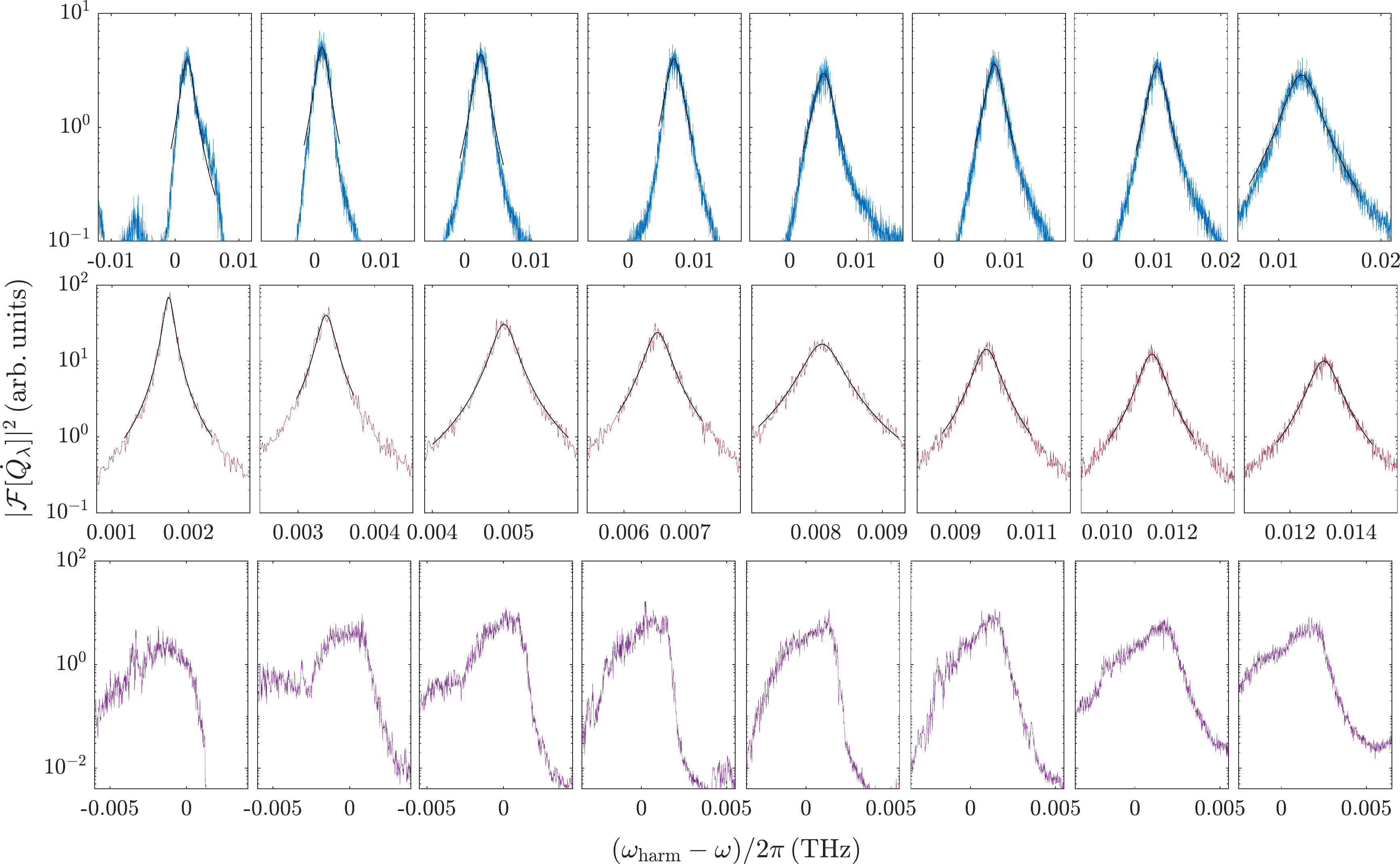}
\caption{\label{fig:P6} Normal mode analysis of CNT (3,3). Spectral energy density of the LA (top row), TW (middle row) and FA (bottom row) normal mode coordinates at $T=300\,\rm{K}$. Columns are arranged in order of increasing wavenumber, $q=(2\pi/400a)\,m$ with $m=(1,\dots,8)$, from left to right. The frequency axis is normalized with mode frequencies $\omega_{\rm{harm}}$ obtained from a harmonic lattice calculation in the static limit. Black lines in the top and middle row show a Lorentzian fit used to determine phonon lifetimes $\tau_{\lambda}^{\rm{MD}}$ as per Eq.~(14) of the main article.} 
\end{figure*}

\begin{figure*}[ht]
\centering
\includegraphics[width=0.95\columnwidth]{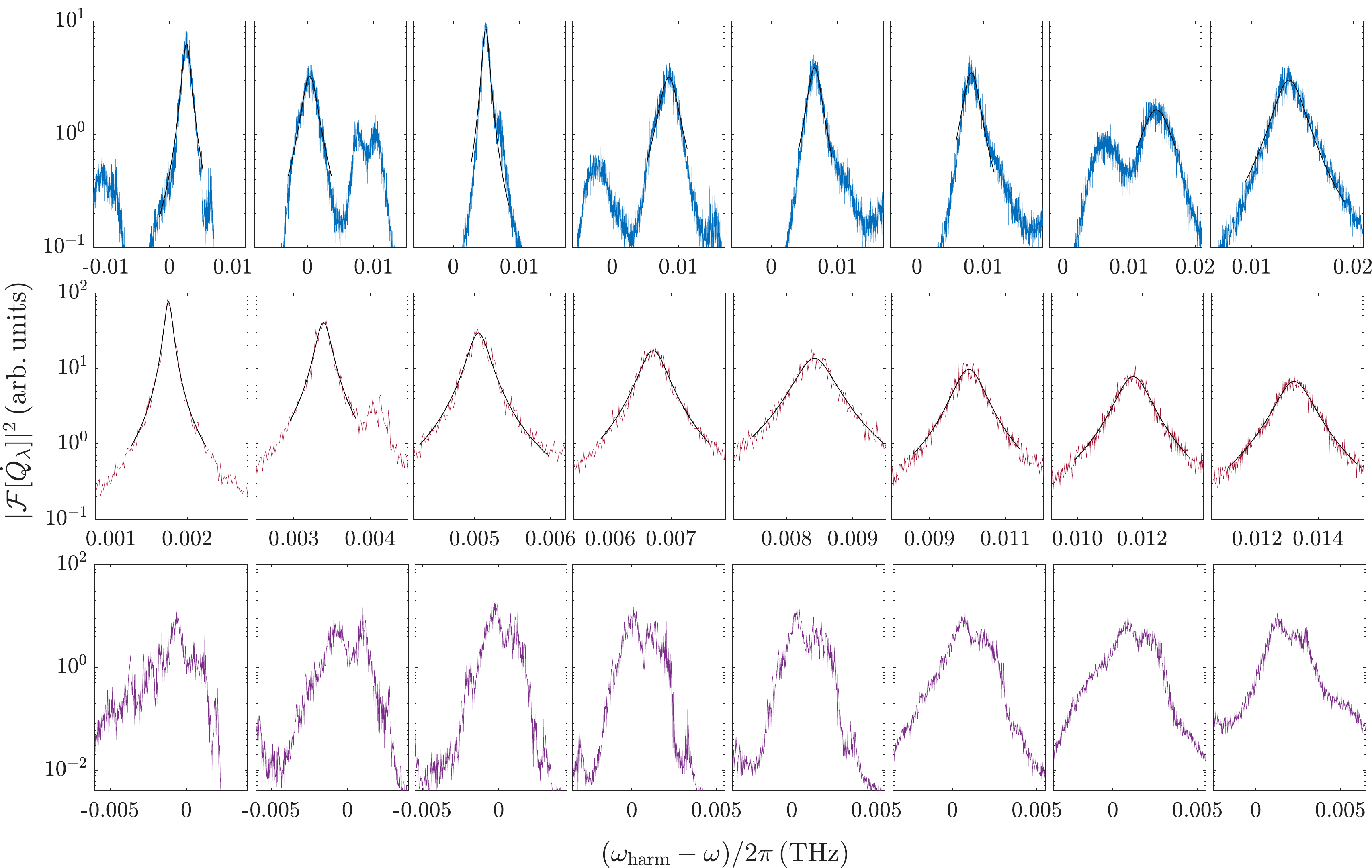}
\caption{\label{fig:P7} Same as Fig.~\ref{fig:P6}, but for CNT (4,4).}
\end{figure*}

\begin{figure*}[ht]
\centering
\includegraphics[width=0.95\columnwidth]{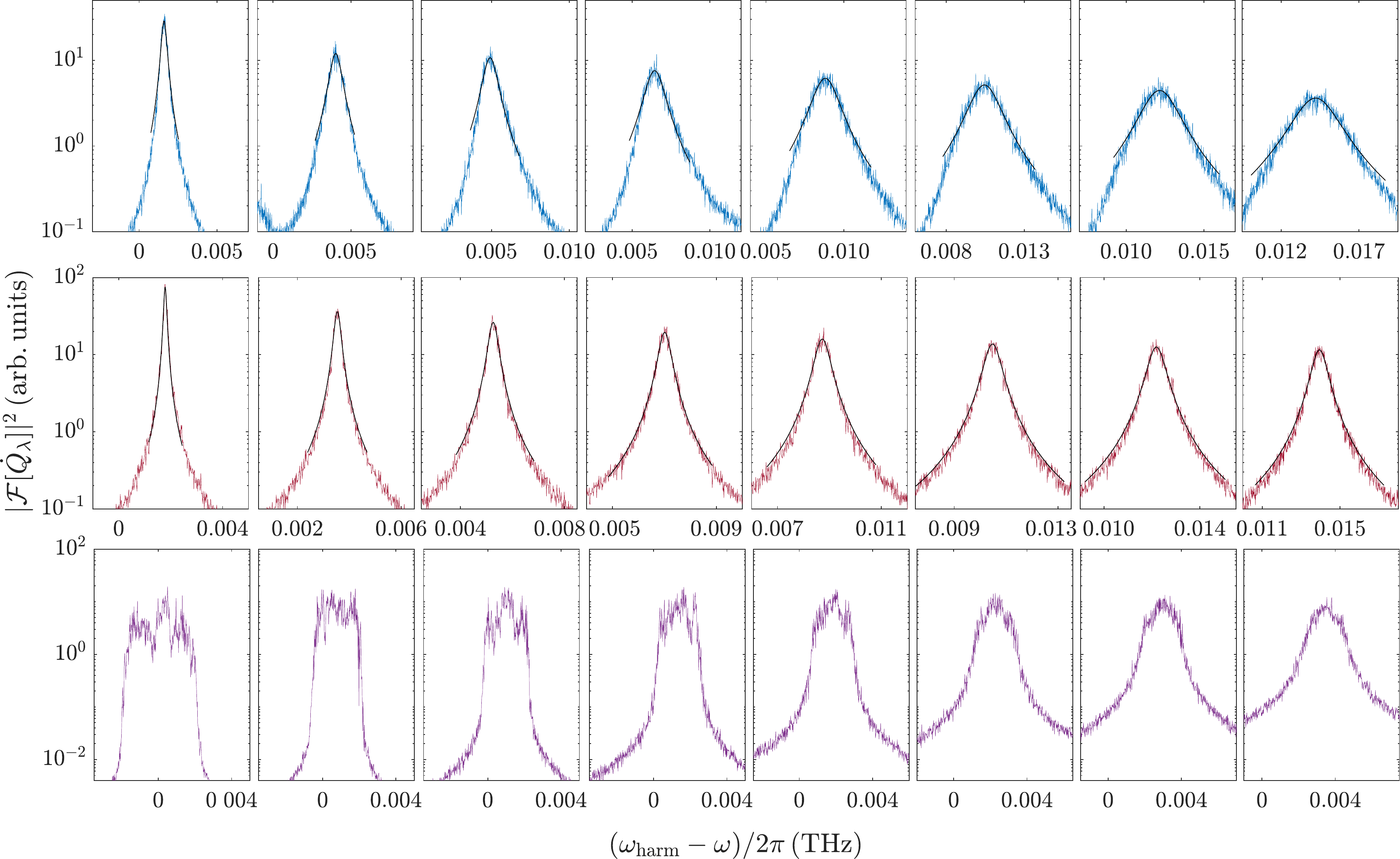}
\caption{\label{fig:P8} Same as Fig.~\ref{fig:P6}, but for CNT (7,7).} 
\end{figure*}

\begin{figure*}[ht]
\centering
\includegraphics[width=0.95\columnwidth]{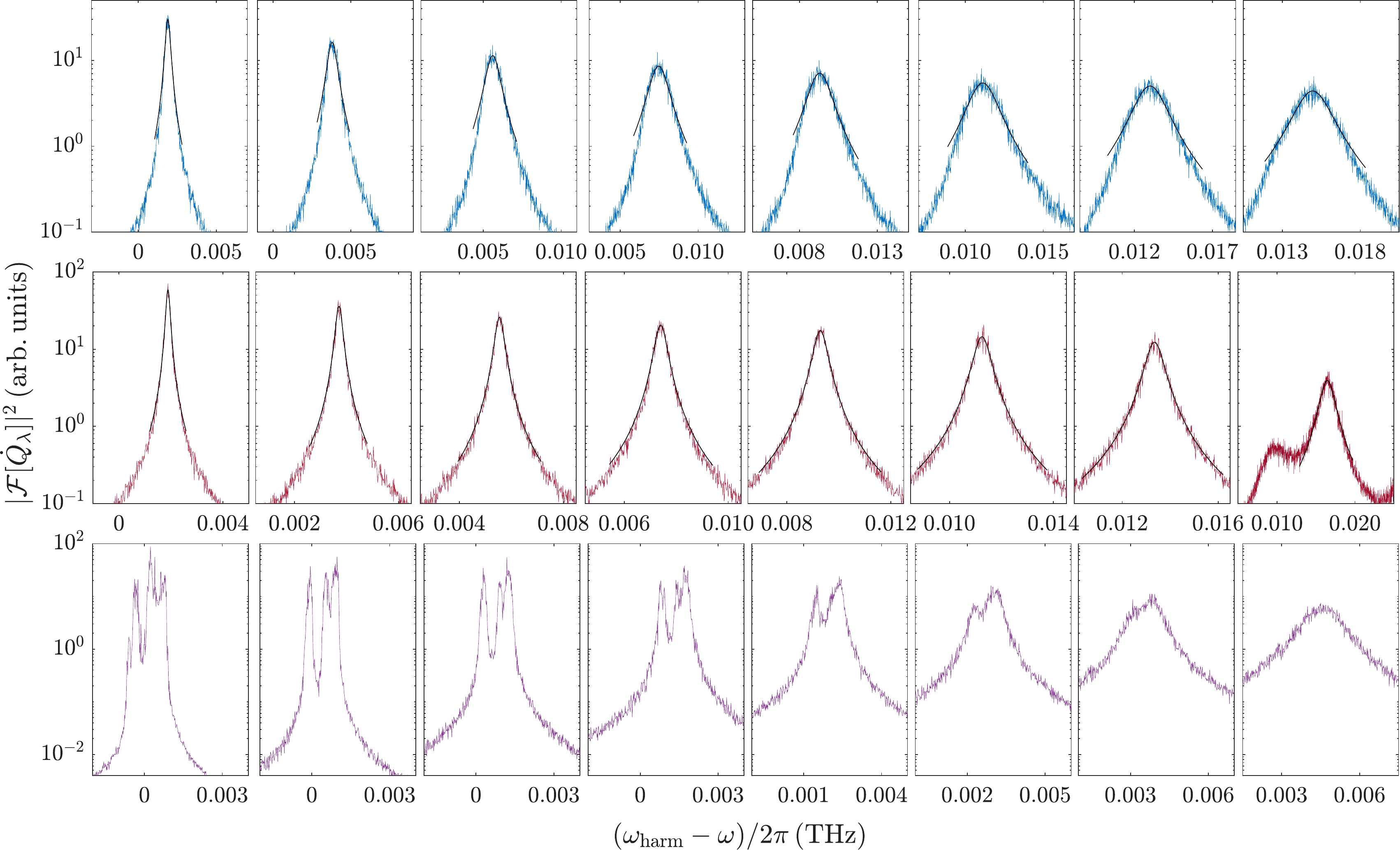}
\caption{\label{fig:P9} Same as Fig.~\ref{fig:P6}, but for CNT (10,10).} 
\end{figure*}

For CNTs of different diameter, each comprising $N=400$ translational unit cells, we present the spectral energy densities of long wavelength longitudinal (LA), twisting (TW) and flexural (FA) modes in Figs.~\ref{fig:P6}-\ref{fig:P9}. According to expectations, the two degenerate FA modes of each tube exhibit similar spectra such that only one out of the two FA mode spectra is shown.


\bibliography{SI}